\documentclass[twocolumn, superscriptaddress]{revtex4-1}
\usepackage{amsmath}
\usepackage{booktabs}
\usepackage[utf8]{inputenc}
\usepackage{xcolor}
\usepackage{graphicx}
\usepackage{hyperref}
\usepackage{xspace}
\usepackage{orcidlink}
\usepackage{lineno}
\usepackage{tabularx}
\usepackage{multirow}
\usepackage{array}
\usepackage{acronym}

\newcommand{\tabref}[1]{Tab.~\ref{#1}}
\renewcommand{\eqref}[1]{Eq.~(\ref{#1})}

\newcommand{\figref}[1]{Fig.~\ref{#1}}
\newcommand{\refref}[1]{Ref.~#1}
\newcommand{\secref}[1]{Sec.~\ref{#1}}

\begin{document}

\title{New Methods for Offline GstLAL Analyses}

\author{Prathamesh Joshi \orcidlink{0000-0002-4148-4932}}
\email{prathamesh.joshi@ligo.org}
\affiliation{Department of Physics, The Pennsylvania State University, University Park, PA 16802, USA}
\affiliation{Institute for Gravitation and the Cosmos, The Pennsylvania State University, University Park, PA 16802, USA}
\affiliation{School of Physics, Georgia Institute of Technology, Atlanta, GA 30332, USA}

\author{Leo Tsukada  \orcidlink{0000-0003-0596-5648}}
\email{leo.tsukada@ligo.org}
\affiliation{Department of Physics, The Pennsylvania State University, University Park, PA 16802, USA}
\affiliation{Institute for Gravitation and the Cosmos, The Pennsylvania State University, University Park, PA 16802, USA}
\affiliation{Department of Physics and Astronomy, University of Nevada, Las Vegas, 4505 South Maryland Parkway, Las Vegas, NV 89154, USA}
\affiliation{Nevada Center for Astrophysics, University of Nevada, Las Vegas, NV 89154, USA}

\author{Chad Hanna}
\affiliation{Department of Physics, The Pennsylvania State University, University Park, PA 16802, USA}
\affiliation{Institute for Gravitation and the Cosmos, The Pennsylvania State University, University Park, PA 16802, USA}
\affiliation{Department of Astronomy and Astrophysics, The Pennsylvania State University, University Park, PA 16802, USA}
\affiliation{Institute for Computational and Data Sciences, The Pennsylvania State University, University Park, PA 16802, USA}

\author{Shomik Adhicary \orcidlink{0009-0004-2101-5428}}
\affiliation{Department of Physics, The Pennsylvania State University, University Park, PA 16802, USA}
\affiliation{Institute for Gravitation and the Cosmos, The Pennsylvania State University, University Park, PA 16802, USA}

\author{Debnandini Mukherjee  \orcidlink{0000-0001-7335-9418}}
\affiliation{NASA Marshall Space Flight Center, Huntsville, AL 35811, USA}
\affiliation{Center for Space Plasma and Aeronomic Research, University of Alabama in Huntsville, Huntsville, AL 35899, USA}

\author{Wanting Niu \orcidlink{0000-0003-1470-532X}}
\affiliation{Department of Physics, The Pennsylvania State University, University Park, PA 16802, USA}
\affiliation{Institute for Gravitation and the Cosmos, The Pennsylvania State University, University Park, PA 16802, USA}

\author{Shio Sakon \orcidlink{0000-0002-5861-3024}}
\affiliation{Department of Physics, The Pennsylvania State University, University Park, PA 16802, USA}
\affiliation{Institute for Gravitation and the Cosmos, The Pennsylvania State University, University Park, PA 16802, USA}

\author{Divya Singh \orcidlink{0000-0001-9675-4584}}
\affiliation{Department of Physics, The Pennsylvania State University, University Park, PA 16802, USA}
\affiliation{Institute for Gravitation and the Cosmos, The Pennsylvania State University, University Park, PA 16802, USA}
\affiliation{Department of Physics, University of California, Berkeley, CA 94720, USA}

\author{Pratyusava Baral \orcidlink{0000-0001-6308-211X}}
\affiliation{Leonard E.\ Parker Center for Gravitation, Cosmology, and Astrophysics, University of Wisconsin-Milwaukee, Milwaukee, WI 53201, USA}

\author{Amanda Baylor \orcidlink{0000-0003-0918-0864}}
\affiliation{Leonard E.\ Parker Center for Gravitation, Cosmology, and Astrophysics, University of Wisconsin-Milwaukee, Milwaukee, WI 53201, USA}

\author{Kipp Cannon \orcidlink{0000-0003-4068-6572}}
\affiliation{RESCEU, The University of Tokyo, Tokyo, 113-0033, Japan}

\author{Sarah Caudill}
\affiliation{Department of Physics, University of Massachusetts, Dartmouth, MA 02747, USA}
\affiliation{Center for Scientific Computing and Data Science Research, University of Massachusetts, Dartmouth, MA 02747, USA}

\author{Bryce Cousins \orcidlink{0000-0002-7026-1340}}
\affiliation{Department of Physics, University of Illinois, Urbana, IL 61801 USA}
\affiliation{Department of Physics, The Pennsylvania State University, University Park, PA 16802, USA}
\affiliation{Institute for Gravitation and the Cosmos, The Pennsylvania State University, University Park, PA 16802, USA}

\author{Jolien D. E. Creighton \orcidlink{0000-0003-3600-2406}}
\affiliation{Leonard E.\ Parker Center for Gravitation, Cosmology, and Astrophysics, University of Wisconsin-Milwaukee, Milwaukee, WI 53201, USA}

\author{Becca Ewing}
\affiliation{Department of Physics, The Pennsylvania State University, University Park, PA 16802, USA}
\affiliation{Institute for Gravitation and the Cosmos, The Pennsylvania State University, University Park, PA 16802, USA}

\author{Heather Fong}
\affiliation{Department of Physics and Astronomy, University of British Columbia, Vancouver, BC, V6T 1Z4, Canada}
\affiliation{RESCEU, The University of Tokyo, Tokyo, 113-0033, Japan}
\affiliation{Graduate School of Science, The University of Tokyo, Tokyo 113-0033, Japan}

\author{Richard N. George \orcidlink{0000-0002-7797-7683}}
\affiliation{Center for Gravitational Physics, University of Texas at Austin, Austin, TX 78712, USA}

\author{Patrick Godwin \orcidlink{0000-0002-7489-4751}}
\affiliation{LIGO Laboratory, California Institute of Technology, MS 100-36, Pasadena, California 91125, USA}
\affiliation{Department of Physics, The Pennsylvania State University, University Park, PA 16802, USA}
\affiliation{Institute for Gravitation and the Cosmos, The Pennsylvania State University, University Park, PA 16802, USA}

\author{Reiko Harada}
\affiliation{RESCEU, The University of Tokyo, Tokyo, 113-0033, Japan}
\affiliation{Graduate School of Science, The University of Tokyo, Tokyo 113-0033, Japan}

\author{Yun-Jing Huang \orcidlink{0000-0002-2952-8429}}
\affiliation{Department of Physics, The Pennsylvania State University, University Park, PA 16802, USA}
\affiliation{Institute for Gravitation and the Cosmos, The Pennsylvania State University, University Park, PA 16802, USA}

\author{Rachael Huxford}
\affiliation{Minnesota Supercomputing Institute, University of Minnesota, Minneapolis, MN 55455, USA}

\author{James Kennington \orcidlink{0000-0002-6899-3833}}
\affiliation{Department of Physics, The Pennsylvania State University, University Park, PA 16802, USA}
\affiliation{Institute for Gravitation and the Cosmos, The Pennsylvania State University, University Park, PA 16802, USA}

\author{Soichiro Kuwahara}
\affiliation{RESCEU, The University of Tokyo, Tokyo, 113-0033, Japan}
\affiliation{Graduate School of Science, The University of Tokyo, Tokyo 113-0033, Japan}

\author{Alvin K. Y. Li \orcidlink{0000-0001-6728-6523}}
\affiliation{RESCEU, The University of Tokyo, Tokyo, 113-0033, Japan}
\affiliation{Graduate School of Science, The University of Tokyo, Tokyo 113-0033, Japan}

\author{Ryan Magee \orcidlink{0000-0001-9769-531X}}
\affiliation{LIGO Laboratory, California Institute of Technology, Pasadena, CA 91125, USA}

\author{Duncan Meacher \orcidlink{0000-0001-5882-0368}}
\affiliation{Leonard E.\ Parker Center for Gravitation, Cosmology, and Astrophysics, University of Wisconsin-Milwaukee, Milwaukee, WI 53201, USA}

\author{Cody Messick \orcidlink{0000-0002-8230-3309}}
\affiliation{Leonard E.\ Parker Center for Gravitation, Cosmology, and Astrophysics, University of Wisconsin-Milwaukee, Milwaukee, WI 53201, USA}

\author{Soichiro Morisaki \orcidlink{0000-0002-8445-6747}}
\affiliation{Institute for Cosmic Ray Research, The University of Tokyo, 5-1-5 Kashiwanoha, Kashiwa, Chiba 277-8582, Japan}

\author{Alexander Pace \orcidlink{0009-0003-4044-0334}}
\affiliation{Department of Physics, The Pennsylvania State University, University Park, PA 16802, USA}
\affiliation{Institute for Gravitation and the Cosmos, The Pennsylvania State University, University Park, PA 16802, USA}

\author{Cort Posnansky \orcidlink{0009-0009-7137-9795}}
\affiliation{Department of Physics, The Pennsylvania State University, University Park, PA 16802, USA}
\affiliation{Institute for Gravitation and the Cosmos, The Pennsylvania State University, University Park, PA 16802, USA}

\author{Anarya Ray \orcidlink{0000-0002-7322-4748}}
\affiliation{Leonard E.\ Parker Center for Gravitation, Cosmology, and Astrophysics, University of Wisconsin-Milwaukee, Milwaukee, WI 53201, USA}
\affiliation{Center for Interdischiplinary Exploration and Research in Astrophysics, Northwestern University, IL 60201, USA}

\author{Surabhi Sachdev \orcidlink{0000-0002-0525-2317}}
\affiliation{School of Physics, Georgia Institute of Technology, Atlanta, GA 30332, USA}
\affiliation{Leonard E.\ Parker Center for Gravitation, Cosmology, and Astrophysics, University of Wisconsin-Milwaukee, Milwaukee, WI 53201, USA}

\author{Stefano Schmidt \orcidlink{0000-0002-8206-8089}}
\affiliation{Nikhef, Science Park 105, 1098 XG, Amsterdam, The Netherlands}
\affiliation{Institute for Gravitational and Subatomic Physics (GRASP), Utrecht University, Princetonplein 1, 3584 CC Utrecht, The Netherlands}

\author{Urja Shah \orcidlink{0000-0001-8249-7425}}
\affiliation{School of Physics, Georgia Institute of Technology, Atlanta, GA 30332, USA}

\author{Ron Tapia}
\affiliation{Department of Physics, The Pennsylvania State University, University Park, PA 16802, USA}
\affiliation{Institute for Computational and Data Sciences, The Pennsylvania State University, University Park, PA 16802, USA}

\author{Koh Ueno \orcidlink{0000-0003-3227-6055}}
\affiliation{RESCEU, The University of Tokyo, Tokyo, 113-0033, Japan}

\author{Aaron Viets \orcidlink{0000-0002-4241-1428}}
\affiliation{Concordia University Wisconsin, Mequon, WI 53097, USA}

\author{Leslie Wade}
\affiliation{Department of Physics, Hayes Hall, Kenyon College, Gambier, Ohio 43022, USA}

\author{Madeline Wade \orcidlink{0000-0002-5703-4469}}
\affiliation{Department of Physics, Hayes Hall, Kenyon College, Gambier, Ohio 43022, USA}

\author{Zach Yarbrough \orcidlink{0000-0002-9825-1136}}
\affiliation{Department of Physics and Astronomy, Louisiana State University, Baton Rouge, LA 70803, USA}

\author{Noah Zhang \orcidlink{0009-0003-3361-5538}}
\affiliation{School of Physics, Georgia Institute of Technology, Atlanta, GA 30332, USA}

\date{\today}


\begin{abstract}
In this work, we present new methods implemented in the GstLAL offline gravitational wave search.
These include a technique to reuse the matched filtering data products from a GstLAL
online analysis, which hugely reduces the time and computational resources required 
to obtain offline results; a technique to combine these results with a separate
search for heavier black hole mergers, enabling detections from a larger set of 
gravitational wave sources; changes to the likelihood ratio which increases the sensitivity
of the analysis; and two separate changes to the background estimation, allowing
more precise significance estimation of gravitational wave candidates. Some of these
methods increase the sensitivity of the analysis, whereas others correct previous
mis-estimations of sensitivity by eliminating false positives.
These methods have been adopted for GstLAL's offline results during the fourth
observing run of LIGO, Virgo, and KAGRA (O4). To test these new methods, we perform
an offline analysis over one chunk of O3 data, lasting from May 12 19:36:42 UTC 2019
to May 21 14:45:08 UTC 2019, and compare it with previous GstLAL results over the same
period of time. We show that cumulatively these methods afford around a 50\% - 100\% increase
in sensitivity in the highest mass space, while simultaneously increasing the
reliability of results, and making them more reusable and computationally cheaper.

\end{abstract}

\maketitle

\acrodef{LSC}[LSC]{LIGO Scientific Collaboration}
\acrodef{LVC}[LVC]{LIGO Scientific and Virgo Collaboration}
\acrodef{LVK}[LVK]{LIGO Scientific, Virgo and KAGRA Collaborations}
\acrodef{aLIGO}{Advanced Laser Interferometer Gravitational-Wave Observatory}
\acrodef{aVirgo}{Advanced Virgo}
\acrodef{LIGO}[LIGO]{Laser Interferometer Gravitational-Wave Observatory}
\acrodef{IFO}[IFO]{interferometer}
\acrodef{LHO}[LHO]{LIGO-Hanford}
\acrodef{LLO}[LLO]{LIGO-Livingston}
\acrodef{O2}[O2]{second observing run}
\acrodef{O1}[O1]{first observing run}
\acrodef{O3}[O3]{third observing run}
\acrodef{O3a}[O3a]{first half of the third observing run}
\acrodef{O3b}[O3b]{second half of the third observing run}
\acrodef{O4a}[O4a]{first part of the fourth observing run}
\acrodef{O4b}[O4b]{second part of the fourth observing run}
\acrodef{O4}[O4]{fourth observing run}

\acrodef{SSM}[SSM]{subsolar-mass}
\acrodef{BH}[BH]{black hole}
\acrodef{BBH}[BBH]{binary black hole}
\acrodef{BNS}[BNS]{binary neutron star}
\acrodef{IMBH}[IMBH]{intermediate-mass black hole}
\acrodef{NS}[NS]{neutron star}
\acrodef{BHNS}[BHNS]{black hole--neutron star binaries}
\acrodef{NSBH}[NSBH]{neutron star--black hole binary}
\acrodef{PBH}[PBH]{primordial black hole binaries}
\acrodef{CBC}[CBC]{compact binary coalescence}
\acrodef{GW}[GW]{gravitational wave}
\acrodef{GWH}[GW]{gravitational-wave}
\acrodef{DBH}[DBH]{dispasstive black hole binaries}
\acrodef{SVD}[SVD]{singular value decomposition}
\acrodef{SNR}[SNR]{signal-to-noise ratio}
\acrodef{LR}[LR]{likelihood ratio}
\acrodef{KDE}[KDE]{kernel density estimate}
\acrodef{FAR}[FAR]{false alarm rate}
\acrodef{PSD}[PSD]{power spectral density}
\acrodef{GR}[GR]{general relativity}
\acrodef{NR}[NR]{numerical relativity}
\acrodef{PN}[PN]{post-Newtonian}
\acrodef{EOB}[EOB]{effective-one-body}
\acrodef{ROM}[ROM]{reduced-order model}
\acrodef{IMR}[IMR]{inspiral--merger--ringdown}
\acrodef{EOS}[EoS]{equation of state}
\acrodef{FF}[FF]{fitting factor}
\acrodef{FT}[FT]{Fourier Transform}

\acrodef{LAL}[LAL]{LIGO Algorithm Library}
\acrodef{GWTC}[GWTC]{Gravitational Wave Transient Catalog}

\newcommand{\PN}[0]{\ac{PN}\xspace}
\newcommand{\BBH}[0]{\ac{BBH}\xspace}
\newcommand{\BNS}[0]{\ac{BNS}\xspace}
\newcommand{\BH}[0]{\ac{BH}\xspace}
\newcommand{\NR}[0]{\ac{NR}\xspace}
\newcommand{\GW}[0]{\ac{GW}\xspace}
\newcommand{\SNR}[0]{\ac{SNR}\xspace}
\newcommand{\SSM}[0]{\ac{SSM}\xspace}
\newcommand{\aLIGO}[0]{\ac{aLIGO}\xspace}
\newcommand{\PSD}[0]{\ac{PSD}\xspace}
\newcommand{\GR}[0]{\ac{GR}\xspace}
\newcommand{\EOS}[0]{\ac{EOS}\xspace}
\newcommand{\LVC}[0]{\ac{LVC}\xspace}


\newcommand{\GSTLAL}{GstLAL\xspace}
\newcommand{\IMRPHENOMD}{IMRPhenomD\xspace}
\newcommand{\MANIFOLD}{{\fontfamily{qcr}\selectfont manifold}\xspace}
\newcommand{\SBANK}{{\fontfamily{qcr}\selectfont SBank}\xspace}

\newcommand\hmm[1]{\ifnum\ifhmode\spacefactor\else2000\fi>1500 \uppercase{#1}\else#1\fi}

\newcommand{\IGWNALERT}{\texttt{igwn-alert}\xspace}

\newcommand{\MDCSTART}{5 Jan. 2020 15:59:42}
\newcommand{\MDCEND}{14 Feb. 2020 15:59:42}

\newcommand{\TOTALTEMPLATES}{\ensuremath{1.8 \times 10^6}}
\newcommand{\CHECKERBOARDTEMPLATES}{\ensuremath{9 \times 10^5}}
\newcommand{\NUMSVDBANKS}{\ensuremath{\sim 1000}}
\newcommand{\TEMPLATESPERSUBBANK}{\ensuremath{\sim 500}}
\newcommand{\NUMSUBBANKSPERSVD}{\ensuremath{2}}
\newcommand{\SVDTOLERANCE}{\ensuremath{99.999\%}}
\newcommand{\PSDFFTLENGTH}{\ensuremath{4~\mathrm{seconds}}}
\newcommand{\FRAMELENGTH}{\ensuremath{1~\mathrm{second}}}
\newcommand{\BUFFERBLOCKSIZE}{\ensuremath{4096~\mathrm{bytes}}}
\newcommand{\FIRSTRIDE}{\ensuremath{0.25~\mathrm{seconds}}}
\newcommand{\TRIGGERSNRTHRESHOLD}{\ensuremath{4.0}}
\newcommand{\COINCTHRESHOLD}{\ensuremath{0.005~\mathrm{seconds}}}
\newcommand{\HTGATETHRESHOLDMIN}{\ensuremath{15.0}}
\newcommand{\HTGATETHRESHOLDMAX}{\ensuremath{100.0}}
\newcommand{\HTGATEMCHIRPMIN}{\ensuremath{0.8}}
\newcommand{\HTGATEMCHIRPMAX}{\ensuremath{45.0}}
\newcommand{\HTGATEMIN}{\ensuremath{\sim 15}}
\newcommand{\HTGATEMAX}{\ensuremath{\sim 325}}
\newcommand{\LRSNAPSHOT}{\ensuremath{4}}
\newcommand{\LRCOMPRESSION}{\ensuremath{0.003}}
\newcommand{\FARTRIALSFACTOR}{\ensuremath{2}}
\newcommand{\UPLOADCADENCE}{\ensuremath{4}}
\newcommand{\UPLOADCADENCEMDCTWELVE}{\ensuremath{2}}
\newcommand{\UPLOADDT}{\ensuremath{0.2}}
\newcommand{\SINGLESPENALTYMDCELEVEN}{\ensuremath{12}}
\newcommand{\SINGLESPENALTYOFOUR}{\ensuremath{13}}
\newcommand{\XISQMISMATCHRANGE}{\ensuremath{0.1-10\%}}

\newcommand{\TESTSUITECOINCWINDOW}{\ensuremath{\pm 1}}
\newcommand{\INJSNRFLOW}{\ensuremath{10.0}}
\newcommand{\INJSNRFHI}{\ensuremath{1600.0}}

\newcommand{\VT}{\ensuremath{\langle VT \rangle}}
\newcommand{\SPINZ}{\ensuremath{s_{i,z}}}
\newcommand{\CHIP}{\ensuremath{\chi_p}}
\newcommand{\MCHIRP}{\ensuremath{\mathcal{M}_c}\xspace}
\newcommand{\CHIEFF}{\ensuremath{\chi_{\mathrm{eff}}}}
\newcommand{\PASTRO}{\ensuremath{p(\mathrm{astro})}}
\newcommand{\MSUN}{\ensuremath{M_{\odot}}}
\newcommand{\TEND}{\ensuremath{t_{\mathrm{end}}}}

\newcommand{\TOTALINJECTIONS}{$5\times10^4$}
\newcommand{\BNSMAXZ}{\ensuremath{0.15}}
\newcommand{\NSBHMAXZ}{\ensuremath{0.25}}
\newcommand{\BBHMAXZ}{\ensuremath{1.9}}
\newcommand{\MDCDURATION}{\ensuremath{3.456\times10^6}}
\newcommand{\INJECTIONSPACING}{\ensuremath{\sim40}}

\newcommand{\PASTROTHRESHOLD}{\ensuremath{0.50}}

\newcommand{\DECISIVESNRTHRESH}{\ensuremath{8.0}}
\newcommand{\NETWORKSNRTHRESH}{\ensuremath{10.0}}

\newcommand{\ALLABOVEDECSNRTHRESH}{\ensuremath{1522}}
\newcommand{\ALLINBANKABOVEDECSNRTHRESH}{\ensuremath{1457}}
\newcommand{\BBHINBANKABOVEDECSNRTHRESH}{\ensuremath{597}}
\newcommand{\BNSINBANKABOVEDECSNRTHRESH}{\ensuremath{482}}
\newcommand{\NSBHINBANKABOVEDECSNRTHRESH}{\ensuremath{378}}

\newcommand{\HIGHFARTHRESH}{$1$ per hour}
\newcommand{\LOWFARTHRESH}{$2$ per day}
\newcommand{\ONEPERHOUR}{\ensuremath{2.78\times10^{-4}~\mathrm{Hz}}}
\newcommand{\TWOPERDAY}{\ensuremath{2.31\times10^{-5}~\mathrm{Hz}}}
\newcommand{\ONEPERMONTH}{\ensuremath{3.85\times10^{-7}~\mathrm{Hz}}}
\newcommand{\TWOPERYEAR}{\ensuremath{3.16\times10^{-8}~\mathrm{Hz}}}

\newcommand{\ALLINBANKEFFICIENCY}[1]{%
	\IfEqCase{#1}{%
		{ONEPERHOUR}{\ensuremath{0.87}}%
		{TWOPERDAY}{\ensuremath{0.84}}%
		{ONEPERMONTH}{\ensuremath{0.78}}%
		{TWOPERYEAR}{\ensuremath{0.74}}%
	}[\PackageError{ALLINBANKEFFICIENCY}{Undefined option: #1}{}]
}%

\newcommand{\BBHINBANKEFFICIENCY}[1]{%
	\IfEqCase{#1}{%
		{ONEPERHOUR}{\ensuremath{0.87}}%
		{TWOPERDAY}{\ensuremath{0.84}}%
		{ONEPERMONTH}{\ensuremath{0.77}}%
		{TWOPERYEAR}{\ensuremath{0.71}}%
	}[\PackageError{BBHINBANKEFFICIENCY}{Undefined option: #1}{}]
}%

\newcommand{\BNSINBANKEFFICIENCY}[1]{%
	\IfEqCase{#1}{%
		{ONEPERHOUR}{\ensuremath{0.95}}%
		{TWOPERDAY}{\ensuremath{0.95}}%
		{ONEPERMONTH}{\ensuremath{0.89}}%
		{TWOPERYEAR}{\ensuremath{0.86}}%
	}[\PackageError{BNSINBANKEFFICIENCY}{Undefined option: #1}{}]
}%

\newcommand{\NSBHINBANKEFFICIENCY}[1]{%
	\IfEqCase{#1}{%
		{ONEPERHOUR}{\ensuremath{0.77}}%
		{TWOPERDAY}{\ensuremath{0.71}}%
		{ONEPERMONTH}{\ensuremath{0.65}}%
		{TWOPERYEAR}{\ensuremath{0.62}}%
	}[\PackageError{NSBHINBANKEFFICIENCY}{Undefined option: #1}{}]
}%

\newcommand{\ALLABOVEDECSNRTHRESHMDCTWELVE}{\ensuremath{653}}
\newcommand{\ALLINBANKABOVEDECSNRTHRESHMDCTWELVE}{\ensuremath{621}}
\newcommand{\BBHINBANKABOVEDECSNRTHRESHMDCTWELVE}{\ensuremath{243}}
\newcommand{\BNSINBANKABOVEDECSNRTHRESHMDCTWELVE}{\ensuremath{209}}
\newcommand{\NSBHINBANKABOVEDECSNRTHRESHMDCTWELVE}{\ensuremath{169}}

\newcommand{\ALLINBANKEFFICIENCYMDCTWELVE}[1]{%
	\IfEqCase{#1}{%
		{ONEPERHOUR}{\ensuremath{0.88}}%
		{TWOPERDAY}{\ensuremath{0.86}}%
		{ONEPERMONTH}{\ensuremath{0.83}}%
		{TWOPERYEAR}{\ensuremath{0.81}}%
	}[\PackageError{ALLINBANKEFFICIENCYMDCTWELVE}{Undefined option: #1}{}]
}%

\newcommand{\BBHINBANKEFFICIENCYMDCTWELVE}[1]{%
	\IfEqCase{#1}{%
		{ONEPERHOUR}{\ensuremath{0.92}}%
		{TWOPERDAY}{\ensuremath{0.90}}%
		{ONEPERMONTH}{\ensuremath{0.87}}%
		{TWOPERYEAR}{\ensuremath{0.86}}%
	}[\PackageError{BBHINBANKEFFICIENCYMDCTWELVE}{Undefined option: #1}{}]
}%

\newcommand{\BNSINBANKEFFICIENCYMDCTWELVE}[1]{%
	\IfEqCase{#1}{%
		{ONEPERHOUR}{\ensuremath{0.95}}%
		{TWOPERDAY}{\ensuremath{0.93}}%
		{ONEPERMONTH}{\ensuremath{0.91}}%
		{TWOPERYEAR}{\ensuremath{0.88}}%
	}[\PackageError{BNSINBANKEFFICIENCYMDCTWELVE}{Undefined option: #1}{}]
}%

\newcommand{\NSBHINBANKEFFICIENCYMDCTWELVE}[1]{%
	\IfEqCase{#1}{%
		{ONEPERHOUR}{\ensuremath{0.74}}%
		{TWOPERDAY}{\ensuremath{0.72}}%
		{ONEPERMONTH}{\ensuremath{0.68}}%
		{TWOPERYEAR}{\ensuremath{0.66}}%
	}[\PackageError{NSBHINBANKEFFICIENCYMDCTWELVE}{Undefined option: #1}{}]
}%

\newcommand{\MEAN}[1]{%
	\IfEqCase{#1}{%
		{MASSRATIO}{\ensuremath{1.39}}%
		{MCHIRP}{\ensuremath{0.15}}%
		{SPIN1Z}{\ensuremath{7.27}}%
		{SPIN2Z}{\ensuremath{2.81}}%
		{CHIEFF}{\ensuremath{5.77}}%
		{ENDTIME}{\ensuremath{6.23}}%
	}[\PackageError{MEAN}{Undefined option: #1}{}]
}%

\newcommand{\STDEV}[1]{%
	\IfEqCase{#1}{%
		{MASSRATIO}{\ensuremath{2.86}}%
		{MCHIRP}{\ensuremath{0.45}}%
		{SPIN1Z}{\ensuremath{285}}%
		{SPIN2Z}{\ensuremath{183}}%
		{CHIEFF}{\ensuremath{252}}%
		{ENDTIME}{\ensuremath{30.22}}%
	}[\PackageError{STDEV}{Undefined option: #1}{}]
}%

\newcommand{\BNSMCHIRPMEAN}{\ensuremath{2.06\times10^{-4}}}
\newcommand{\BNSMCHIRPSTDEV}{\ensuremath{8.33\times10^{-4}}}

\newcommand{\NSBHMCHIRPMEAN}{\ensuremath{-2.14\times10^{-4}}}
\newcommand{\NSBHMCHIRPSTDEV}{\ensuremath{6.26\times10^{-3}}}

\newcommand{\BBHMCHIRPMEAN}{\ensuremath{1.54\times10^{-1}}}
\newcommand{\BBHMCHIRPSTDEV}{\ensuremath{4.53\times10^{-1}}}

\newcommand{\BNSENDTIMEMEAN}{\ensuremath{-0.90}}
\newcommand{\BNSENDTIMESTDEV}{\ensuremath{18.0}}

\newcommand{\NSBHENDTIMEMEAN}{\ensuremath{18.7}}
\newcommand{\NSBHENDTIMESTDEV}{\ensuremath{59.3}}

\newcommand{\BBHENDTIMEMEAN}{\ensuremath{6.03}}
\newcommand{\BBHENDTIMESTDEV}{\ensuremath{11.3}}

\newcommand{\QFIFTY}[1]{%
	\IfEqCase{#1}{%
		{MASSRATIO}{\ensuremath{0.45}}%
		{MCHIRP}{\ensuremath{0.007}}%
		{SPIN1Z}{\ensuremath{1.24}}%
		{SPIN2Z}{\ensuremath{1.72}}%
		{CHIEFF}{\ensuremath{1.34}}%
		{ENDTIME}{\ensuremath{3.8}}%
	}[\PackageError{QFIFTY}{Undefined option: #1}{}]
}%

\newcommand{\QSEVENTYFIVE}[1]{%
	\IfEqCase{#1}{%
		{MASSRATIO}{\ensuremath{1.67}}%
		{MCHIRP}{\ensuremath{0.33}}%
		{SPIN1Z}{\ensuremath{3.82}}%
		{SPIN2Z}{\ensuremath{5.33}}%
		{CHIEFF}{\ensuremath{3.71}}%
		{ENDTIME}{\ensuremath{9.78}}%
	}[\PackageError{QSEVENTYFIVE}{Undefined option: #1}{}]
}%

\newcommand{\QNINETY}[1]{%
	\IfEqCase{#1}{%
		{MASSRATIO}{\ensuremath{4.97}}%
		{MCHIRP}{\ensuremath{0.73}}%
		{SPIN1Z}{\ensuremath{13.8}}%
		{SPIN2Z}{\ensuremath{17.5}}%
		{CHIEFF}{\ensuremath{10.8}}%
		{ENDTIME}{\ensuremath{25.6}}%
	}[\PackageError{QNINETY}{Undefined option: #1}{}]
}%

\newcommand{\GPCYRS}{\ensuremath{\mathrm{Gpc}^3\mathrm{yrs}}}
\newcommand{\INJECTEDVT}[1]{%
	\IfEqCase{#1}{%
		{BNS}{\ensuremath{1.08\times10^{-1}}}%
		{NSBH}{\ensuremath{4.34\times10^{-1}}}%
		{BBH}{\ensuremath{29.1}}%
	}[\PackageError{INJECTEDVT}{Undefined option: #1}{}]
}%

\newcommand{\VTTWOPERDAY}[1]{%
	\IfEqCase{#1}{%
		{BNS}{\ensuremath{3.49\times10^{-4}}}%
		{NSBH}{\ensuremath{8.08\times10^{-4}}}%
		{BBH}{\ensuremath{1.23\times10^{-1}}}%
	}[\PackageError{VTTWOPERDAY}{Undefined option: #1}{}]
}%

\newcommand{\VTNETSNR}[1]{%
	\IfEqCase{#1}{%
		{BNS}{\ensuremath{4.41\times10^{-4}}}%
		{NSBH}{\ensuremath{1.59\times10^{-3}}}%
		{BBH}{\ensuremath{1.52\times10^{-1}}}%
	}[\PackageError{VTNETSNR}{Undefined option: #1}{}]
}%

\newcommand{\VTDECSNR}[1]{%
	\IfEqCase{#1}{%
		{BNS}{\ensuremath{1.47\times10^{-4}}}%
		{NSBH}{\ensuremath{4.98\times10^{-3}}}%
		{BBH}{\ensuremath{5.46\times10^{-2}}}%
	}[\PackageError{VTDECSNR}{Undefined option: #1}{}]
}%


\newcommand{\SEARCHEDAREAQFIFTY}[1]{%
	\IfEqCase{#1}{%
		{ALL}{\ensuremath{271}}%
		{TRIPLE}{\ensuremath{31.9}}%
		{DOUBLE}{\ensuremath{301}}%
		{SINGLE}{\ensuremath{3150}}%
	}[\PackageError{SEARCHEDAREAQFIFTY}{Undefined option: #1}{}]
}%

\newcommand{\SEARCHEDAREAQSEVENTYFIVE}[1]{%
	\IfEqCase{#1}{%
		{ALL}{\ensuremath{1080}}%
		{TRIPLE}{\ensuremath{140}}%
		{DOUBLE}{\ensuremath{893}}%
		{SINGLE}{\ensuremath{10,400}}%
	}[\PackageError{SEARCHEDAREAQSEVENTYFIVE}{Undefined option: #1}{}]
}%

\newcommand{\SEARCHEDAREAQNINETY}[1]{%
	\IfEqCase{#1}{%
		{ALL}{\ensuremath{3910}}%
		{TRIPLE}{\ensuremath{357}}%
		{DOUBLE}{\ensuremath{2470}}%
		{SINGLE}{\ensuremath{18,400}}%
	}[\PackageError{SEARCHEDAREAQNINETY}{Undefined option: #1}{}]
}%

\newcommand{\SEARCHEDPROBQFIFTY}[1]{%
	\IfEqCase{#1}{%
		{ALL}{\ensuremath{0.53}}%
		{TRIPLE}{\ensuremath{0.58}}%
		{DOUBLE}{\ensuremath{0.52}}%
		{SINGLE}{\ensuremath{0.59}}%
	}[\PackageError{SEARCHEDPROBQFIFTY}{Undefined option: #1}{}]
}%

\newcommand{\SEARCHEDPROBQSEVENTYFIVE}[1]{%
	\IfEqCase{#1}{%
		{ALL}{\ensuremath{0.79}}%
		{TRIPLE}{\ensuremath{0.84}}%
		{DOUBLE}{\ensuremath{0.77}}%
		{SINGLE}{\ensuremath{0.78}}%
	}[\PackageError{SEARCHEDPROBQSEVENTYFIVE}{Undefined option: #1}{}]
}%

\newcommand{\SEARCHEDPROBQNINETY}[1]{%
	\IfEqCase{#1}{%
		{ALL}{\ensuremath{0.93}}%
		{TRIPLE}{\ensuremath{0.96}}%
		{DOUBLE}{\ensuremath{0.92}}%
		{SINGLE}{\ensuremath{0.92}}%
	}[\PackageError{SEARCHEDPROBQNINETY}{Undefined option: #1}{}]
}%

\newcommand{\BNSTOBNS}{\ensuremath{90.3\%}}
\newcommand{\BNSTONSBH}{\ensuremath{9.7\%}}

\newcommand{\NSBHTONSBH}{\ensuremath{64.1\%}}
\newcommand{\NSBHTOBBH}{\ensuremath{33.8\%}}
\newcommand{\NSBHTOBNS}{\ensuremath{2.10\%}}

\newcommand{\BBHTOBBH}{\ensuremath{100\%}}

\newcommand{\TERRTOTERR}{\ensuremath{2.60\%}}
\newcommand{\TERRTOBBH}{\ensuremath{68.8\%}}
\newcommand{\TERRTONSBH}{\ensuremath{19.5\%}}
\newcommand{\TERRTOBNS}{\ensuremath{9.10\%}}

\newcommand{\BNSTOBNSMDCTWELVE}{\ensuremath{79.8\%}}
\newcommand{\BNSTONSBHMDCTWELVE}{\ensuremath{20.2\%}}

\newcommand{\NSBHTONSBHMDCTWELVE}{\ensuremath{92.1\%}}
\newcommand{\NSBHTOBBHMDCTWELVE}{\ensuremath{6.83\%}}
\newcommand{\NSBHTOBNSMDCTWELVE}{\ensuremath{1.02\%}}

\newcommand{\BBHTOBBHMDCTWELVE}{\ensuremath{99.5\%}}
\newcommand{\BBHTONSBHMDCTWELVE}{\ensuremath{0.05\%}}

\newcommand{\TERRTOBBHMDCTWELVE}{\ensuremath{76.2\%}}
\newcommand{\TERRTONSBHMDCTWELVE}{\ensuremath{23.8\%}}

\newcommand{\OTHREEOPA}{\ensuremath{1.2}} 

\newcommand{\MDCGWIFOS}[1]{%
	\IfEqCase{#1}{%
		{GW200112}{L1}%
		{GW200115}{H1L1}%
		{GW200128}{H1L1}%
		{GW200129}{H1L1V1}%
		{GW200202}{H1L1}%
		{GW200208q}{H1L1}%
		{GW200208am}{H1L1}%
		{GW200209}{H1L1}%
		{GW200210}{H1L1}%
	}[\PackageError{MDCGWIFOS}{Undefined option: #1}{}]
}%

\newcommand{\MDCGWSNR}[1]{%
	\IfEqCase{#1}{%
		{GW200112}{\ensuremath{18.46}}%
		{GW200115}{\ensuremath{11.48}}%
		{GW200128}{\ensuremath{9.98}}%
		{GW200129}{\ensuremath{26.30}}%
		{GW200202}{\ensuremath{11.09}}%
		{GW200208q}{\ensuremath{10.56}}%
		{GW200208am}{\ensuremath{8.00}}%
		{GW200209}{\ensuremath{9.96}}%
		{GW200210}{\ensuremath{9.28}}%
	}[\PackageError{MDCGWSNR}{Undefined option: #1}{}]
}%

\newcommand{\MDCGWFAR}[1]{%
	\IfEqCase{#1}{%
		{GW200112}{\ensuremath{1.01\times10^{-7}}}%
		{GW200115}{\ensuremath{2.55\times10^{-4}}}%
		{GW200128}{\ensuremath{1.44\times10^{-4}}}%
		{GW200129}{\ensuremath{1.78\times10^{-17}}}%
		{GW200202}{\ensuremath{1.69\times10^{-2}}}%
		{GW200208q}{\ensuremath{4.92\times10^{-5}}}%
		{GW200208am}{\ensuremath{2.02\times10^{3}}}%
		{GW200209}{\ensuremath{1.20}}%
		{GW200210}{\ensuremath{3.64\times10^{3}}}%
	}[\PackageError{MDCGWFAR}{Undefined option: #1}{}]
}%

\newcommand{\MDCGWPASTRO}[1]{%
	\IfEqCase{#1}{%
		{GW200112}{\ensuremath{>0.99}}%
		{GW200115}{\ensuremath{>0.99}}%
		{GW200128}{\ensuremath{>0.99}}%
		{GW200129}{\ensuremath{>0.99}}%
		{GW200202}{\ensuremath{>0.99}}%
		{GW200208q}{\ensuremath{>0.99}}%
		{GW200208am}{\ensuremath{0.48}}%
		{GW200209}{\ensuremath{>0.99}}%
		{GW200210}{0.27}%
	}[\PackageError{MDCGWPASTRO}{Undefined option: #1}{}]
}%

\newcommand{\MDCGWMCHIRP}[1]{%
	\IfEqCase{#1}{%
		{GW200112}{\ensuremath{33.37~M_{\odot}}}%
		{GW200115}{\ensuremath{2.58~M_{\odot}}}%
		{GW200128}{\ensuremath{50.74~M_{\odot}}}%
		{GW200129}{\ensuremath{30.66~M_{\odot}}}%
		{GW200202}{\ensuremath{8.15~M_{\odot}}}%
		{GW200208q}{\ensuremath{34.50~M_{\odot}}}%
		{GW200208am}{\ensuremath{66.59~M_{\odot}}}%
		{GW200209}{\ensuremath{39.45~M_{\odot}}}%
		{GW200210}{\ensuremath{7.89~M_{\odot}}}%
	}[\PackageError{MDCGWMCHIRP}{Undefined option: #1}{}]
}%

\newcommand{\OTHREEGWIFOS}[1]{%
	\IfEqCase{#1}{%
		{GW200112}{L1}%
		{GW200115}{H1L1}%
		{GW200128}{--}%
		{GW200129}{H1L1V1}%
		{GW200202}{--}%
		{GW200208q}{--}%
		{GW200208am}{--}%
		{GW200209}{--}%
		{GW200210}{--}%
	}[\PackageError{OTHREEGWIFOS}{Undefined option: #1}{}]
}%

\newcommand{\OTHREEGWSNR}[1]{%
	\IfEqCase{#1}{%
		{GW200112}{\ensuremath{18.79}}%
		{GW200115}{\ensuremath{11.42}}%
		{GW200128}{--}%
		{GW200129}{\ensuremath{26.61}}%
		{GW200202}{--}%
		{GW200208q}{--}%
		{GW200208am}{--}%
		{GW200209}{--}%
		{GW200210}{--}%
	}[\PackageError{OTHREEGWSNR}{Undefined option: #1}{}]
}%

\newcommand{\OTHREEGWFAR}[1]{%
	\IfEqCase{#1}{%
		{GW200112}{\ensuremath{4.05\times10^{-4}}}%
		{GW200115}{\ensuremath{6.61\times10^{-4}}}%
		{GW200128}{\ensuremath{> \OTHREEOPA{}}}%
		{GW200129}{\ensuremath{2.11\times10^{-24}}}%
		{GW200202}{\ensuremath{> \OTHREEOPA{}}}%
		{GW200208q}{--}%
		{GW200208am}{--}%
		{GW200209}{--}%
		{GW200210}{--}%
	}[\PackageError{OTHREEGWFAR}{Undefined option: #1}{}]
}%

\newcommand{\OTHREEGWPASTRO}[1]{%
	\IfEqCase{#1}{%
		{GW200112}{\ensuremath{>0.99}}%
		{GW200115}{\ensuremath{>0.99}}%
		{GW200128}{--}%
		{GW200129}{\ensuremath{>0.99}}%
		{GW200202}{--}%
		{GW200208q}{--}%
		{GW200208am}{--}%
		{GW200209}{--}%
		{GW200210}{--}%
	}[\PackageError{OTHREEGWPASTRO}{Undefined option: #1}{}]
}%

\newcommand{\OTHREEGWMCHIRP}[1]{%
	\IfEqCase{#1}{%
		{GW200112}{\ensuremath{35.37~M_{\odot}}}%
		{GW200115}{\ensuremath{2.57~M_{\odot}}}%
		{GW200128}{--}%
		{GW200129}{\ensuremath{32.74~M_{\odot}}}%
		{GW200202}{--}%
		{GW200208q}{--}%
		{GW200208am}{--}%
		{GW200209}{--}%
		{GW200210}{--}%
	}[\PackageError{OTHREEGWMCHIRP}{Undefined option: #1}{}]
}

\newcommand{\OTHREERETRACTIONS}{23}
\newcommand{\OTHREEGSTLALRETRACTIONS}{15}

\newcommand{\RETRACTIONFAR}{\ensuremath{1.67~\mathrm{per}~\mathrm{year}}}
\newcommand{\RETRACTIONSNR}{\ensuremath{14.5}}
\newcommand{\MDCRETRACTIONFARTHRESH}{one per year}

\newcommand{\BANKMASSLOW}{\ensuremath{1.0~M_{\odot}}}
\newcommand{\BANKMASSHIGH}{\ensuremath{200~M_{\odot}}}

\newcommand{\BHMASSLOW}{\ensuremath{3.0~M_{\odot}}}
\newcommand{\NSMASSLOW}{\ensuremath{1.0~M_{\odot}}}
\newcommand{\NSMASSHIGH}{\ensuremath{3.0 M_{\odot}}}
\newcommand{\TOTALMASSHIGH}{\ensuremath{400.0~M_{\odot}}}
\newcommand{\MASSRATIOHIGH}{\ensuremath{20}}

\newcommand{\NSSPIN}{\ensuremath{0.05}}
\newcommand{\BHSPIN}{\ensuremath{0.99}}
\newcommand{\CHIPBOUND}{\ensuremath{1\times10^{-3}}}

\newcommand{\MCHIRPBOUNDARY}{\ensuremath{1.73~M_{\odot}}}
\newcommand{\LOWMCHIRPWAVEFORM}{\texttt{TaylorF2}}
\newcommand{\HIGHMCHIRPWAVEFORM}{\texttt{SEOBNRv4}}

\newcommand{\MDCELEVENLOFARLATENCY}{\ensuremath{14.58}}
\newcommand{\MDCELEVENHIFARLATENCY}{\ensuremath{10.30}}

\newcommand{\MDCTWELVELOFARLATENCY}{\ensuremath{12.04}}
\newcommand{\MDCTWELVEHIFARLATENCY}{\ensuremath{9.86}}

\section{Introduction}
\label{sec:introduction}

By detecting \acp{GW} from the merger of compact objects like black holes and neutron stars,
the \ac{LVK} have revolutionized the field of gravitational wave astronomy. GW150914 was the first
such detection of a \ac{GW}~\cite{LIGOScientific:2016aoc}, and since then close to 250 \acp{GW} have
been detected~\cite{detectionplot}. While individual detections can often yield important scientific results, e.g., GW150914~\cite{PhysRevLett.116.221101}, GW170817~\cite{PhysRevLett.121.161101, PhysRevLett.123.011102},
GW200105 and GW200115~\cite{LIGOScientific:2021qlt},
even richer scientific implication can be extracted from a collection of \ac{GW} detections. To this end,
the \ac{LVK} maintains a \textit{catalog} of transient \ac{GW} detections, called the \ac{GWTC}.
To date, four versions of the catalog have been released: GWTC-1~\cite{LIGOScientific:2018mvr}, GWTC-2~\cite{gwtc-2},
GWTC-2.1~\cite{gwtc-2.1}, and GWTC-3~\cite{LIGOScientific:2021djp}. These help inform results related to
testing the theory of general relativity~\cite{PhysRevD.100.104036, slaa143, abbott2021tests, EZQUIAGA2021136665},
cosmology~\cite{ligo2017gravitational, Abbott_2021hc}, black hole properties and formation mechnisms~\cite{staa1176, Wang2021, stab2438},
and binary population and merger rates~\cite{Abbott_2019, Abbott_2021, KAGRA:2021duu}.

These results are enabled by \ac{GW} search pipelines, by detecting \acp{GW} in the data produced by \ac{GW} detectors,
such as the two LIGO detectors~\cite{ligo}, the Virgo detector~\cite{virgo}, and the KAGRA detector~\cite{kagra}.
The contents of the catalog are usually compiled from rigorous high-latency ``offline" \ac{GW} searches
rather than low-latency ``online" ones.
GstLAL~\cite{Messick:2016aqy, Sachdev:2019vvd, Cannon:2020qnf, Hanna:2019ezx}
is one such \ac{GW} search pipeline. Like other modeled \ac{GW} searches, it makes use of a bank of waveform templates
predicted by general relativity. These templates are cross-correlated against the data in a process called
matched filtering to calculate the \ac{SNR}.
GstLAL performs matched filtering in the time domain~\cite{Messick:2016aqy, Sachdev:2019vvd, Cannon:2011vi}
This process is used to identify periods of time possibly containing \ac{GW} signals, which
are called ``triggers". Matched filtering is also used to inform the background data, against which triggers are ranked
to evaluate their significance. A likelihood ratio (LR)~\cite{Cannon:2015gha, Tsukada:2023} is calculated
as the ranking statistic. Triggers with a high LR are called \ac{GW} candidates. The LRs of candidates
are then compared with the LR statistics of background triggers, and using the livetime of the analysis,
a \ac{FAR} is calculated for every candidate. Some of these techniques are used by other \ac{GW}
search pipelines, such as IAS~\cite{ias, Zackay_2021}, MBTA~\cite{mbta, Adams_2016},
PyCBC~\cite{Dal_Canton_2021, Davies_2020, pycbc}, and SPIIR~\cite{spiir, spiir_2017}.

This paper is structured as follows. In \secref{sec:software}, we describe the GstLAL offline
workflow. \secref{sec:methodology} is dedicated to describing the new features introduced
in the GstLAL offline workflow in the lead up to, and during the \ac{O4} of the \ac{LVK}.
\secref{sec:results} contains the results of the tests performed to evaluate these features,
and to compare them against \ac{O3} results.

\section{Overview of the GstLAL Offline Analysis}
\label{sec:software}

The GstLAL offline analysis can be broadly divided into two stages, each of which is further divided
into smaller stages. These are:
\begin{enumerate}
	\item{setup stage}
	\begin{enumerate}
		\item{template bank creation stage}
		\item{\ac{PSD} measurement stage}
		\item{\ac{SVD} of templates and template whitening stage}
	\end{enumerate}

	\item{data processing stage}
	\begin{enumerate}
		\item{matched filtering stage}
		\item{rank stage}
	\end{enumerate}
\end{enumerate}

\subsection{Template bank creation stage}
\label{subsection:allsky_temp_bank_design}
For \ac{O4}, two GstLAL template banks, namely the stellar-mass black hole and \ac{IMBH} banks were generated using \MANIFOLD ~\citep{PhysRevD.108.042003, manifold},
which is a binary-tree approach to template bank generation.
All templates in both banks neglect
eccentricity, higher-order modes, precession and matter effects. They additionally assume
that the spins of each component object are aligned
with an orbital angular momentum of the binary. This is done to limit the dimensionality of the bank,
and hence the number of templates contained in it.
While both of these banks are used to implement all-sky searches, in order to be consistent with internal conventions, in this work we will
refer to the former search as the ``AllSky search", while the latter will continue to be called the IMBH search.

\subsubsection{AllSky template bank}
\label{subsubsection:o3bank}
The AllSky template bank includes templates in the \ac{BNS}, \ac{NSBH}, and \ac{BBH}
parameter spaces~\citep{Ewing:2023}, consisting of $1815963$ templates in total. This bank is used for the GstLAL online analysis, as well as
for the bulk of the GstLAL offline results.
It extends from component masses of $1$ $M_\odot$ to $200$ $M_\odot$. The absolute values of dimensionless spin components
are capped at 0.99 for components of the binary template that are black holes (defined as having a mass above $3 M_\odot$), and 0.05 for components that are neutron stars (defined as having a mass below $3 M_\odot$).
Specifications of the AllSky template bank are shown in ~\tabref{table:allsky_design}, and 
\figref{figure:methodology_allsky_imbh} shows a representation of the AllSky bank in the $m_1$- $m_2$ space.
Readers are referred to Ref.~\cite{Sakon:2022ibh} for more details about the AllSky template
bank.

\begin{table}
\begin{center}
        \begin{tabular}{ l | l }
                \hline
                Parameter & AllSky Template Bank \\
                \hline
                \hline
                Primary mass, $m_1$ & $\in [1.0, 200 M_\odot]$  \\
                Secondary Mass, $m_2$ & $\in [1.0, 200 M_\odot]$  \\
                Mass ratio, $q=m_1/m_2$ & $\in [1, 20]$  \\
                Total mass, $m_1 + m_2$ & $\in [2.0, 400 M_\odot]$  \\
                Dimensionless spin, $s_{\rm {i,z}}$, for $m_i \leq 3.0 M_\odot$ & $\lvert s_{\rm {i,z}} \rvert \leq 0.05$ \\
                Dimensionless spin, $s_{\rm {i,z}}$, for $m_i \geq 3.0 M_\odot$ & $\lvert s_{\rm {i,z}} \rvert \leq 0.99$ \\
                Lower frequency cut-off & 15 Hz \footnote{Due to enforcement of a maximum duration of 128 seconds, some lower mass tempaltes have a higher lower frequency cut-off} \\
                Higher frequency cut-off & 1024 Hz \\
                Waveform approximant & IMRPhenomD \\
                Minimum match & 97 $\%$ \\
                \ac{PSD} & O4 projected \ac{PSD} \footnote{\cite{ligo_psd_dcc}} \\
                \hline
                Total number of templates & 1815963 \\
                \hline
        \end{tabular}
        \caption{Parameters of the \GSTLAL \ac{O4} AllSky template bank.}
        \label{table:allsky_design}
\end{center}
\end{table}

In comparison, the \GSTLAL \ac{O3} AllSky template bank covered component masses
$m_i \in [1.0, 400] M_\odot$.
It also limited templates to spin-aligned systems and
set an upper limit of $0.05$ on the magnitude of the spin for component masses
below $3 M_\odot$.
The spin magnitudes for component masses above $3 M_\odot$ were set to $0.999$.
The \ac{O3} template bank used a minimum match value of 99\% for the \ac{BNS}
space, in contrast to the \ac{O4} value of 97\%.
The parameter choices of the \ac{O3} template bank is shown in Table II of Ref.~\cite{gwtc-2}.
As a result of different template bank generation algorithms and parameter choices,
the \ac{O3} AllSky bank consisted of $1758763$ templates.

The upper mass limit of the \ac{O4} AllSky bank is lower as compared to that of the \ac{O3} one.
This was intentionally done, with the \ac{O4} IMBH bank (explained below) covering this space instead.
The other differences in the \ac{O4} AllSky bank as compared to the \ac{O3} one were motivated by
better template coverage of parameter spaces of interest with the new \MANIFOLD method, while still minimizing
the number of templates required.

\subsubsection{IMBH template bank}
\begin{figure}
    \centering
    \includegraphics[width=\linewidth]{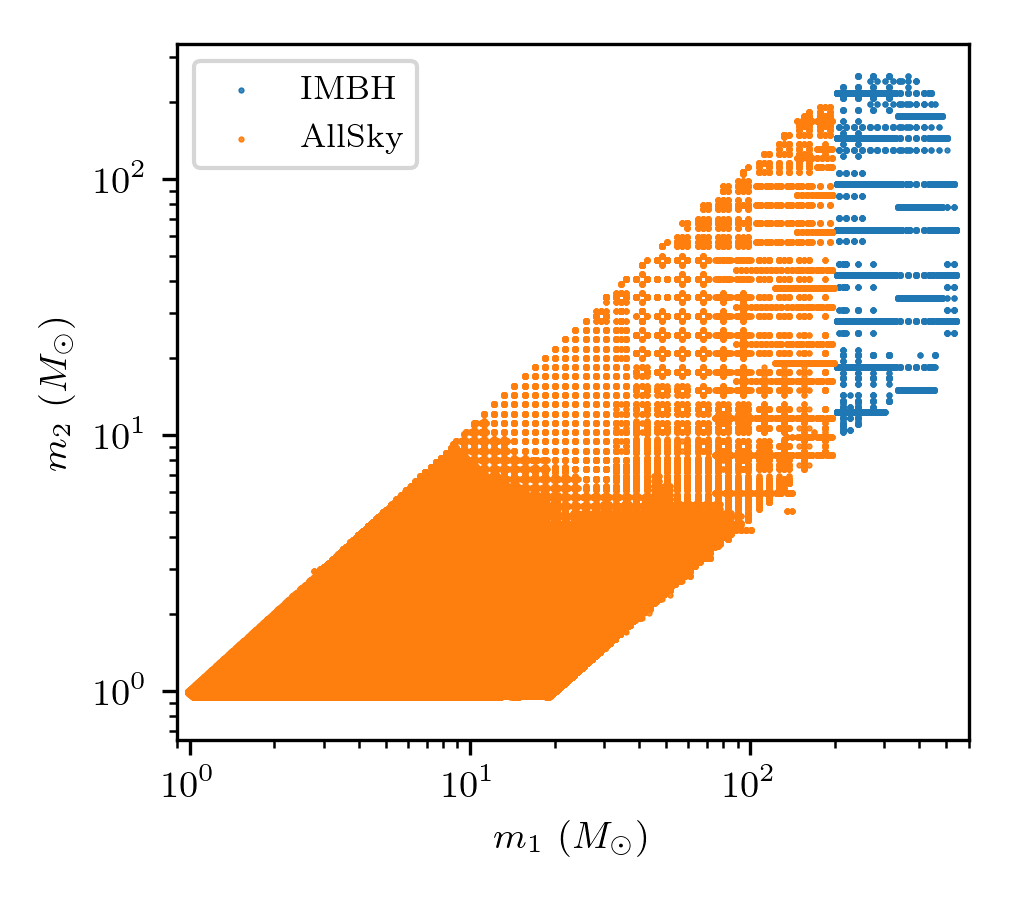}
	\caption{AllSky templates and \ac{IMBH} templates on the $\rm{log} \left(m_1\right)$-$\rm{log} \left(m_2\right)$ plane. Here, the orange dots with $m_1 \leq 200 M_\odot$ are AllSky templates and blue dots with $m_1 \geq 200 M_\odot$ are \ac{IMBH} templates.}
	\label{figure:methodology_allsky_imbh}
\end{figure}

In contrast to the AllSky bank, the \ac{IMBH} bank is much smaller,
and is only used to augment the GstLAL offline results in the \ac{IMBH} parameter space.
The \ac{IMBH} bank covers the mass parameter space higher than what the AllSky 
template bank targets, as shown in Table ~\ref{table:methodology_imbh_param}.
The lower limit of $m_1$ is set to $203$ $M_\odot$ such that the \ac{IMBH} bank 
is an extention to the AllSky bank and the templates of the two template banks 
do not overlap in the $m_1$-$m_2$ space, as shown in \figref{figure:methodology_allsky_imbh}.
The dimensionless spins extend from -0.69 to 0.99. The reason for the lower cutoff (-0.69) to be
higher than that of the AllSky bank (-0.99) is to prevent the ringdown frequency of the templates
from getting close to the lower frequency cutoff for matched filtering used by the IMBH search (10 Hz).
Details of the parameters of the \ac{IMBH} bank are shown in \tabref{table:methodology_imbh_param}. 
\figref{figure:methodology_allsky_imbh} also shows a plot of the \ac{IMBH} bank in the $m_1$-$m_2$ space.

\begin{table}
\begin{center}
	\begin{tabular}{ l | l }
		\hline
		Parameter & IMBH Template Bank \\
		\hline
		\hline
		Primary mass, $m_1$ & $\in [203, 543 M_\odot]$  \\
		Secondary Mass, $m_2$ & $\in [10, 255 M_\odot]$  \\
		Mass ratio, $q=m_1/m_2$ & $\in [1, 20]$  \\
		Total mass, $m_1 + m_2$ & $\in [215, 660 M_\odot]$  \\
		Dimensionless spin, $s_{\rm {i,z}}$ & $s_{\rm {i,z}} \in [-0.69, 0.99]$ \\
		Lower frequency cut-off & 10 Hz \\
		Higher frequency cut-off & 1024 Hz \\
		Waveform approximant & IMRPhenomD \\
		Minimum match & 99 $\%$ \\
		\ac{PSD} & O4 projected \ac{PSD} \footnote{\cite{ligo_psd_dcc}} \\
		\hline
		Total number of templates & 14728 \\
		\hline
	\end{tabular}
	\caption{Parameters of the \GSTLAL \ac{O4} \ac{IMBH} template bank.}
	\label{table:methodology_imbh_param} 
\end{center}
\end{table}

\subsubsection{Population model}
Population models provide weights to templates which represent our prior knowledge
of the astrophysical distribution of \ac{GW} sources~\citep{Fong:2018}. This is used in the \ac{LR} calculation,
as well as to compute probabilities of astrophysical
origin of candidates~\citep{Ray:2023nhx}.
In \ac{O4}, the population models for both the AllSky and \ac{IMBH} banks were
generated using \MANIFOLD. They implement a Salpeter mass function to assign weights to templates, defined as:
\begin{equation}
p(m_1, m_2, s_{1,z}, s_{2,z}) \propto \frac{m_1^{-2.35}}{m_1 - m_\mathrm{min}}
\end{equation}
For both the AllSky and IMBH population models, the value of $m_{min}$ is set to 0.8, which is slightly less
than the minimum mass in the combined AllSky+IMBH template bank. The Salpeter function was chosen as a simple
equation that works well over a large range of masses to approximate the astrophysical distibution of \ac{GW} sources
inferred by previous studies~\cite{Abbott_2019, Abbott_2021, KAGRA:2021duu}.

\subsection{\ac{PSD} measurement stage}
The \ac{PSD} is the frequency representation of detector noise, and is used in the matched filter
calculation. This is done by whitening both the data and the templates with the \ac{PSD} measured
from the data. GstLAL has the ability to measure the \ac{PSD} of the data in real time and dynamically
whiten the data during the matched filtering stage. However, this ability does not extend to template 
whitening, and the templates need to be whitened before the analysis starts. The \ac{PSD} used for this
purpose is measured from the data in the \ac{PSD} measurement stage. The data is first divided into
smaller pieces of up to 8 hours depending on the continuity of segments when each detector is operational.
The \ac{PSD} over each such piece of data is individually measured, and the median over each frequency bin
is taken, which removes
any transient noise features contributing to the \ac{PSD} of a single piece of data. The resultant \ac{PSD} is used to whiten the templates.
More details about \ac{PSD} measurement can be found in Ref.~\cite{Messick:2016aqy}.

\subsection{\Ac{SVD} of templates and template whitening stage}
Next, as discussed in Ref.~\cite{Sakon:2022ibh}, the template bank is sorted by linear
combinations of their Post-Newtonian phase coefficients~\cite{Morisaki:2020oqk} and split into ``template bins", each of around 1000 templates, which are whitened using the median \ac{PSD} described above.
These template bins further go through a process called \ac{SVD} to save computational cost of the matched filtering~\cite{PhysRevD.82.044025}.
The templates in one such template bin are processed together for the purpose of matched filtering and
background estimation, constituting a single matched-filtering job, and hence this template binning allows for a large-scale parallelization in both online and offline workflows.

\subsection{Matched filtering stage}
Each template within a template bin is matched filtered
with the data, producing triggers. Alongside \ac{SNR}, a signal consistency statistic, called $\xi^2$
is also calculated for every trigger~\cite{Messick:2016aqy}.
Triggers with contribution from only one detector (referred to as a ``single detector trigger"), during a time
when multiple detectors were operational (referred to as ``coincident time"), are considered to originate from noise~\cite{Joshi:2023ltf}. This is because
\ac{GW} signals are expected to be correlated across detectors, unlike noise. The \ac{SNR} and 
$\xi^2$ statistics of noise triggers are added to the template bin's common noise background, which
is later one of the inputs used to calculate the \ac{LR} for triggers.

In parallel, the \ac{GW} candidates are extracted from the triggers. This is done by retaining
only those triggers within a template bin that have the highest \ac{SNR} in a 0.2 second window. A single \ac{GW} signal
can create triggers via multiple templates, and with current detector sensitivities, it is highly unlikely for multiple
\ac{GW} signals to occur in the same 0.2 second window~/cite{Relton2021}. Hence, with this ``clustering" over \ac{SNR},
duplicate triggers from a single common signal are removed, reducing the amount of data downstream stages need to process.
The processes of populating the background and generating candidates are done in parallel for every template bin.

\subsection{Rank stage}
In this stage, the significance of candidates is evaluated, and the results of the search are produced.
First, the \ac{LR} is calculated for every candidate forwarded by the matched filtering stage. One of the ingredients in calculating the \ac{LR} is the 
\ac{SNR}-$\xi^2$ background populated by the matched filtering stage. To model a smooth distribution of \ac{SNR}-$\xi^2$ statistics, a \ac{KDE} is applied
to the background, and the resulting probability density is used for \ac{LR} evaluation. This gives us
the probability that the (\ac{SNR}, $\xi^2$) of a given candidate arises from noise. This is one of the many
terms in the \ac{LR} equation. More details on the \ac{LR} calculation can be found in Ref.~\cite{Tsukada:2023}.

Next, the candidate list is pruned again by performing a second round of clustering. This time, it is
performed \textit{across} template bins. The candidate with the highest \ac{LR} in an 8 second window is retained.
Apart from eliminating candidates originating from noise, this ensures that a single \ac{GW} signal
will have at most one candidate originating from it in the entire analysis.

Finally, to convert the \ac{LR} to a \ac{FAR}, we need the \ac{LR} distribution of noise triggers from every template bin~\cite{Cannon:2012zt, Cannon:2015gha}. If we use
the same set of noise triggers as was used to populate the \ac{SNR}-$\xi^2$ background, the \ac{LR} distribution
will not be well defined at higher \ac{LR}s due to lower number of statistics. Additionally, since these
noise triggers were obtained during the livetime of the analysis, the lowest \ac{FAR} would be bound by $1$ per livetime. To solve this problem, GstLAL extrapolates the set of noise triggers from every template bin
by drawing samples of \ac{SNR} and $\xi^2$, and assigning random templates as well as arrival times and phases.
This is a computationally cheap operation, and under the assumption that the template, time and phase distribution
of noise triggers is uniform, this process can effectively extrapolate the \ac{LR} distribution of noise
triggers to high \ac{LR}s, enabling the \ac{FAR} calculation of even the most significant candidates.

However, this process does not work particularly well for low \ac{LR}s. To fix this, we can make one more
assumption, that at low \ac{LR}s, even the candidates originate from noise. Consequently, the low \ac{LR} 
distribution of candidates is the same as the low \ac{LR} distribution of noise triggers, and the former
can be used to inform the latter. The method of doing so is called the ``extinction model". Once the \ac{LR}
distribution of noise triggers is calculated, it can be combined with the livetime of the analysis
to produce a \ac{FAR} for every candidate.

\section{New Methods}
\label{sec:methodology}

\subsection{Online Rank}
\label{subsection:online_rank}
During every observing run, the full data are analyzed in near-real time by the GstLAL low-latency or ``online"
analysis. The various stages of the online analysis, up to the matched filtering stage, are largely similar
to those of the offline analysis. Consequently, during \ac{O4}, a new method has been adopted to outsource the 
GstLAL offline analysis' matched filtering stage to the GstLAL online analysis. This method is called
an ``online rank". A description of the differences between the GstLAL online and offline analyses,
as well as a detailed description of the online rank method is provided in Ref.~\cite{joshi2025timesmatchedfiltergravitational}.

As described in Ref.~\cite{joshi2025timesmatchedfiltergravitational}, the online rank method significantly reduces the computational
cost and time needed for offline results. Matched filtering is the vast majority of the computational
burden of a modeled \ac{GW} search. By eliminating the need to repeat matched filtering a second time
for the offline analysis, this method affords a 50\% reduction in computational cost over the course
of an observing run, and specifically a 95 - 99.8\% reduction in time required to obtain offline results.

\subsection{Dropped Data Refiltering}
\begin{figure*}
\includegraphics[width=\textwidth]{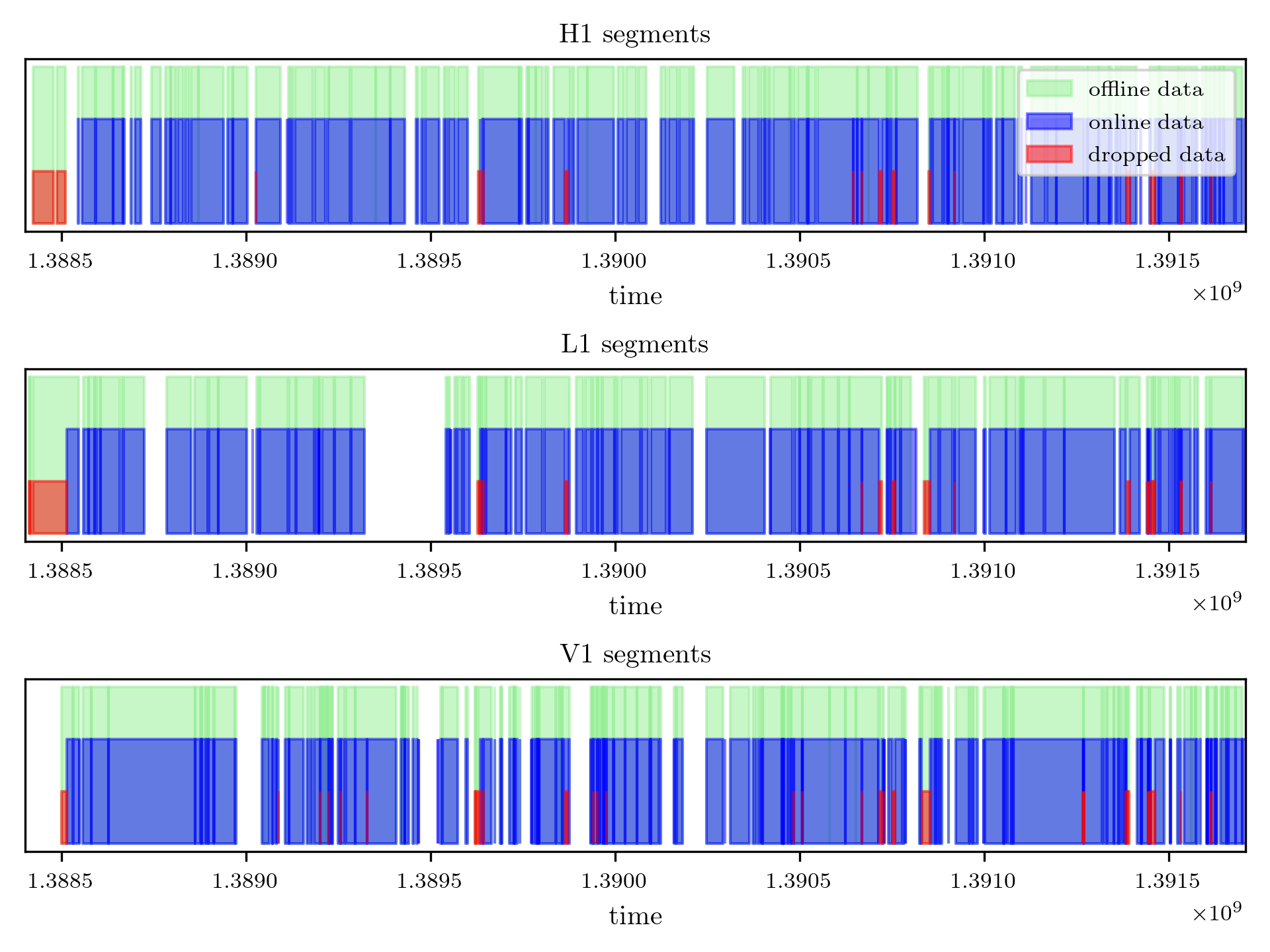}
\caption{\label{fig:dropped_data}
An example of a particular online job's list of dropped data segments.
The dropped data segments at the very start of the plot are because of slightly
different start times of the online and offline analysis, whereas the ones after
that are because of online analysis having failed to analyze those times, either
because the data for those times was dropped in order to keep up with incoming
live data, or because the online job was not functioning at that time. We see that
the amount of dropped data is not too large, but the segments are dispersed
throughout the period of the analysis. Ref.~\cite{joshi2025timesmatchedfiltergravitational} shows that a typical
online analysis drops around 5\% of the total data. This data can be filtered offline
in order to augment the online rank's results. This particular plot was made
using a GstLAL online analysis that participated in the mock data challenge~\cite{mdc_analytics}.
}
\end{figure*}

In order to make the online rank results even more reliable and sensitive, we
can augment the inputs to the online rank with triggers and background data from
times that the online analysis dropped. To do this, for every job in the online
analysis, we take the list of time periods when each job was functional and
producing output.  We then subtract that from the list of times a traditional
offline analysis would have analyzed, which is calculated by external tools as
an extra step in the setup stage of the offline workflow. This leaves us with a
list of times that a particular job has dropped. We calculate such a ``dropped
data segments" for every job. Then, we can set up a traditional offline analysis
using these dropped data segments. The only unusual aspect of this procedure is
that while setting up this ``dropped data refiltering" analysis, each job has
its own segments list rather than a global one for every job. An example of
such a list of dropped data segments is shown in \figref{fig:dropped_data}

By combining the online rank inputs with the results of the dropped data
refiltering analysis, we can be sure the online rank produces offline results
for exactly the same periods of time that the traditional offline analysis would
have. The typical amount of dropped data for any job is around 5\% of the total
time covered by the offline segments~\cite{joshi2025timesmatchedfiltergravitational}.

A point to note is that while we can add data to the online rank using this method,
and we also have the ability to remove data to the granularity of the snapshots
recorded by the online analysis (4 hours, see Ref.~\cite{joshi2025timesmatchedfiltergravitational}), we cannot
remove arbitrarily small amounts of data from an online rank. This means that
we cannot incorporate more refined data quality information, such as small vetoed
periods of time that only become available after the online analysis has finished.

\subsection{\ac{IMBH} Analysis}
The AllSky template bank was designed to be a general-purpose bank for an online analysis.
To obtain more comprehensive offline results, we want to augment the results of the AllSky search
with a search for \ac{IMBH} mergers. The GstLAL \ac{O4} offline results are obtained by performing
a search over the data using the \ac{IMBH} template bank described in \secref{sec:software}, which
is then combined with the AllSky search into a single set of results using the procedue described
in the following subsection.

As shown in Ref.~\cite{Hanna:2019ezx, Tsukada:2023}, the LR contains a term to test the consistency of signals accross the network of detectors,
for observed parameters like SNR and the coalescence time and phase. These parameters can only take on certain possible physical values for a network of detectors and
follow specific correlations that indicate whether a GW trigger is of astrophysical origin.
The correlation among these observables and their joint distribution depend on trigger's template, and hence in principle this signal consistency test should be modified for every template. However, for the purpose of pre-computation efficiency, we adopt a \ac{BNS} template with component masses of
1.4 $M_\odot$, which is shown to be effective \textit{throughout} Allsky template bank. In contrast, the mass range of the \ac{IMBH} template bank is so high that this empirical fact might not be true, and hence for the \ac{IMBH} analysis we use a template with component masses of 60 $M_\odot$.

Other differences in search settings between the AllSky and the IMBH searches include different 
lower frequency cutoffs during matched filtering. AllSky uses 15 Hz, but the 
IMBH search uses 10 Hz as these binaries merge at lower frequencies. The Allsky search uses the 
TaylorF2~\cite{TaylorF2} approximant upto a chirp mass of 1.73 for cost effectivity and the 
SEOBNRv4$\_$ROM~\cite{SEOBNRv4_ROM} beyond that, while the IMBH search only uses the SEOBNRv4$\_$ROM as an 
approximant. 
Note that template placement for both banks was done with the IMRPhenomD approximant~\cite{IMRPhenomD}, but we do not
expect this inconsistency to be problematic.
Similarly, the autocorrelation length used by the Allsky search for calculating the $\xi$
signal-based veto~\cite{Messick:2016aqy} is 701 sample points upto a chirp mass of 1.73, and 351 above that. The IMBH search uses a value of
351 sample points throughout, due to a shorter duration of the waveforms. For this reason, the minimum number of samples
included in a given time slice~\cite{Messick:2016aqy} is 512 for IMBH and 2048 for Allsky. For similar reasons, while splitting
the IMBH bank into template bins, the templates are sorted by their duration which is found to be more effective 
for this part of the parameter space, rather than sorting by linear combinations of their Post-Newtonian phase coefficients.

It is known that the candidates in the \ac{IMBH} search are overwhelmingly single-detector candidates.
This was true even for triggers, before any clustering was done. For example, in the test described in
\secref{sec:results}, out of the total candidates reported by the \ac{IMBH} search, 92.75\% were found to
be single-detector candidates, and out of the set of triggers, 98.57\% were single-detector triggers.
The corresponding numbers for the AllSky search are 3.37\% and 35.07\% respectively. While the root
cause of this is not well understood, the vastly different percentages of single-detector triggers in the 
AllSky and \ac{IMBH} searches indicate that this arises from the matched filtering and coincidence formation
of \ac{IMBH} templates, rather than anything to do with the LR.

\begin{figure}
\includegraphics[width=\columnwidth]{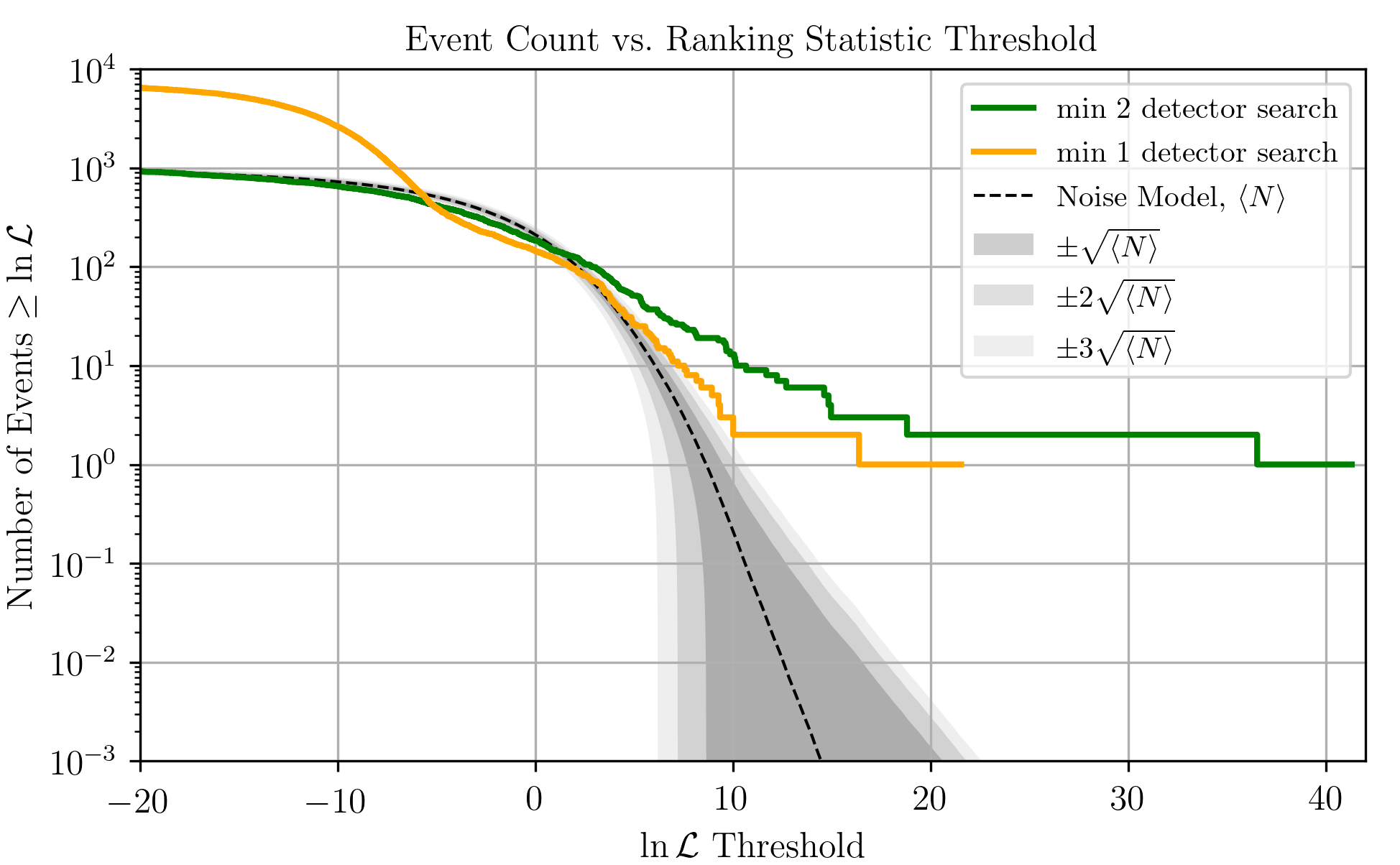}
\caption{\label{fig:imbh_lrs}
This plot shows the results of two \ac{IMBH} searches: one which only considers triggers with two or more
detectors contributing to it, and another with no such restriction. The plot also shows the noise
LR statistics of the former search for reference. We can see that the \ac{LR} statistics of the candidates of the search that only processes
triggers with two or more detectors is well behaved, whereas the other one is not. This is because 
the \ac{IMBH} search, if allowed to process single-detector triggers recovers an overwhelming amount of them, and
the \acp{LR} (and hence FARs) of single-detector triggers are difficult to accurately calculate. As a result, for O4,
GstLAL's \ac{IMBH} search doesn't process single-detector triggers.
}
\end{figure}

The \acp{LR} of single-detector candidates are notoriously difficult to estimate~\cite{callister2017observing}, and
so is the process of \ac{LR} extrapolation of noise triggers. Consequently, the \acp{FAR} of single-detector 
candidates are less reliable than those of multi-detector candidates, and a search with such a high percentage of single-detector candidates is unlikely
to be functional. As a result, for GstLAL's \ac{O4} offline results, the \ac{IMBH} search ignores single
detector triggers, and only processes those with multiple contributing detectors. The effects on the LR
statistics of candidates if single-detector triggers are allowed to be processed are shown in \figref{fig:imbh_lrs}.

\subsection{Combining the AllSky and \ac{IMBH} analyses}
In order to
combine Allsky and \ac{IMBH} searches into a single search, one can naively treat every template bin
from both the AllSky and \ac{IMBH} searches as though they are part of a bigger AllSky+\ac{IMBH} joint template bank.
This means after each template bin assigns \acp{LR} to its candidates, these candidates would be clustered 
together based on their LR, and the noise \ac{LR} statistics of each template bin will be combined via the extinction
model. An implicit assumption of this method is that each template bin's candidate and noise \ac{LR} statistics in
the noise regime are approximately the same.
However, with the different parameter space and settings of the \ac{IMBH} search as compared to AllSky, this
assumption is no longer valid, and hence the naively combined results can be severely biased. To this end, we have developed a new scheme to combine the AllSky and \ac{IMBH} analyses into a single
search.

We first calculate a \textit{weight} for each individual analysis, based on each analysis' contribution to
the final clustered set of candidates. This was done using injection campaigns for both analyses. Injections
are simulated \ac{GW} signals inserted into the data. This weight represents the number of GW signals we expect
each analysis to contribute independent of the other, and is a combination of the analysis' sensitivity, and the 
expected number of GW signals in each parameter space. The weights for the AllSky and \ac{IMBH} analyses
were found to be 0.94 and 0.06. This result was also verified by summing over the population model weights
for the AllSky and \ac{IMBH} templates after taking into consideration the different \ac{SNR} detection thresholds
of the two analyses.

Next, the \acp{FAR} of each analysis' candidates are scaled up by the inverse of their respective weights.
Note that \acp{FAR} are assigned to each analysis separately, and hence the \acp{FAR}
of candidates in the noise regime in both analyses are guaranteed to be distributed similarly as they only depend on the livetime
of the search.
Since the weights sum up to 1, this new set
of candidates is guaranteed to have a consistent distribution of \acp{FAR} in the noise regime, i.e., there
will be on average livetime (expressed in hours) number of candidates with a \ac{FAR} of one per hour or lower, and so on.
Candidates from both analyses are then clustered together by retaining the lowest FAR candidate in an 8 second window.
The \acp{LR} of the candidates are then re-calculated based on their new \acp{FAR}, using the inverse of the mapping
used by the AllSky analysis to convert \acp{LR} into \acp{FAR}. In principle, any such mapping would have worked, and
the AllSky mapping is chosen only for convenience.

\subsection{Modularity and reusability of results}
A general theme of the development work prior to and during \ac{O4} has been about reducing the duplication
of computations done elsewhere. Apart from the online rank described in \secref{subsection:online_rank}, we have made the GstLAL offline
workflow more modular and flexible. In general, offline analyses over large periods of data (such as GstLAL's offline 
results for \ac{O4}) are performed in small chunks, each of around 1 week of data. Alongside the ability
to extract any amount of matched filtering data from the online analysis as specified by the user (to the granularity of
4 hours)~\cite{joshi2025timesmatchedfiltergravitational}, the GstLAL offline workflow also has the ability to use the data from
any set of offline chunks specified by the user. Consequently, one can specify any period of data covered
by any combination of online and offline analyses using any template banks, and they can be combined into a
single result.
Therefore, once a period of data has been matched filtered by a particular template bank via either an
online (complemented by the dropped data refiltering) or offline analysis, it never needs to be filtered using that template bank again. This has greatly
reduced the computational and time burden of GstLAL operations as well as development work during \ac{O4}.

\subsection{Ranking statistic improvements}
\label{subsec:rank_stats}

\begin{figure*}[t]
    \centering
    \begin{minipage}[b]{\columnwidth}
        \includegraphics[width=\columnwidth]{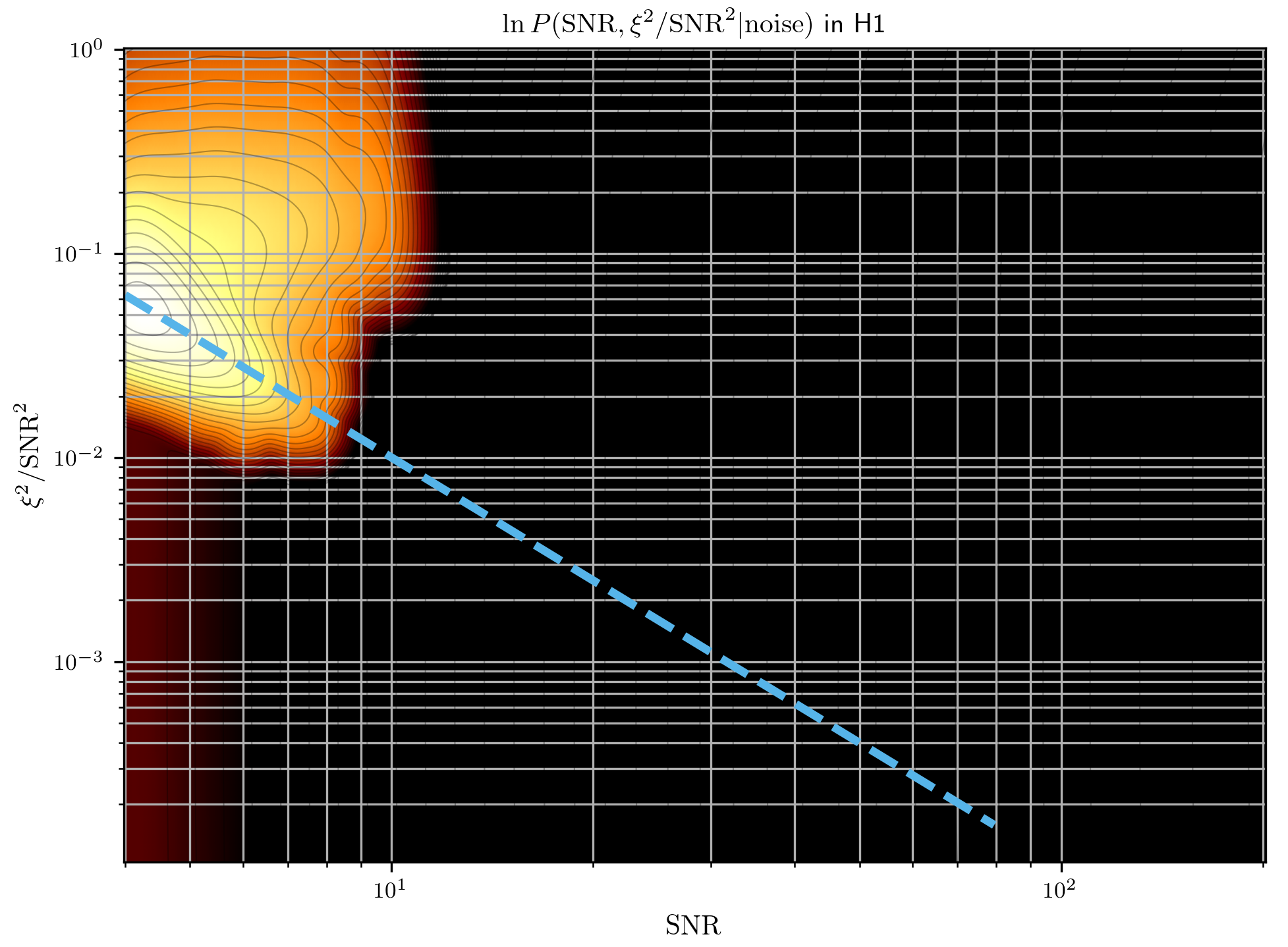}
\caption{\label{fig:bg_plot} Example of $\rho-\xi^2$ noise model for one of the low-mass template groups collected for \ac{LIGO} Hanford detector during the mock data campaign described in~\cite{mdc_analytics} with the lightblue dashed line as a $\xi^2=1$ contour}
    \end{minipage}
    \hfill
    \begin{minipage}[b]{\columnwidth}
        \includegraphics[width=\columnwidth]{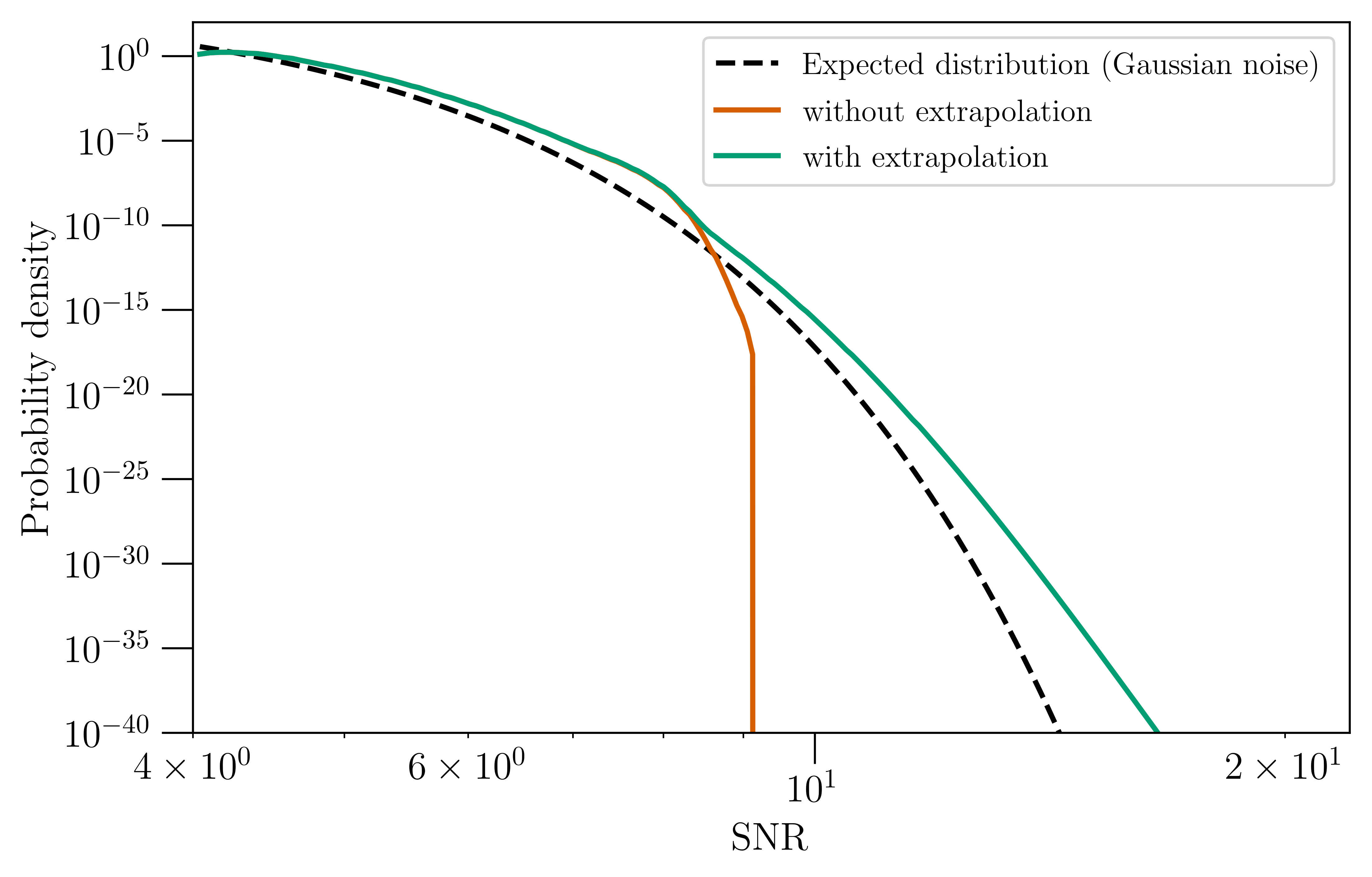}
        \caption{2D $\rho-\xi^2$ noise model sliced at the $\xi^2=1$ contour
        and projected onto the $\rho$ dimension. The noise model with the
        extrapolation closely follows the expected distribution in black curve
        at $\rho>9$ as opposed to the red curve without the extrapolation.}
        \label{fig:snr_pdf}
    \end{minipage}
\end{figure*}

\GSTLAL has adopted the likelihood ratio as the ranking statistic to evaluate
the statistical significance of \ac{GW} candidates~\cite{Cannon:2015gha, Tsukada:2023}.
The likelihood ratio in \GSTLAL takes the form of
\begin{align}
    \label{eq:lr}
    \mathcal{L}=\frac{P\left(\vec{O}, \vec{\rho}, \vec{\xi^2}, \vec{t}, \vec{\phi}, \theta \mid \mathcal{H}_\mathrm{s}\right)}{P\left(\vec{O}, \vec{\rho}, \vec{\xi^2}, \vec{t}, \vec{\phi}, \theta \mid \mathcal{H}_\mathrm{n}\right)},
\end{align}
which represents the probability of obtaining a set of observable parameters, e.g., \ac{SNR} for each detector $\vec{\rho}$,
under the signal hypothesis ($\mathcal{H}_\mathrm{s}$) relative to that under
the noise hypothesis ($\mathcal{H}_\mathrm{n}$).
Here, $\rho$ refers to the SNR of a single detector, and we will use this terminology for the rest of this subsection
to maintain compatibility with others papers on this subject. The improvements in this
ranking statistic prior to \ac{O4}'s start are thoroughly described in Ref.~\cite{Tsukada:2023},
which include the upgraded $\rho-\xi^2$ signal model and removal of signal contamination. In particular, the upgraded $\rho-\xi^2$ signal model achieved $\sim20\%$ increase in the search sensitivity, and hence has led to a major contribution of \GSTLAL to \ac{LVK}'s \ac{GW} detections in \ac{O4} so far.

Apart from the improvements mentioned above, we have modified the $\rho-\xi^2$ noise
model to make more accurate estimate of event's significance in some part of ($\rho,
\xi^2$) parameter space. As described in Ref.~\cite{Tsukada:2023, Joshi:2023ltf},
\GSTLAL collects single-detector triggers during coincident time into the
2D $\rho-\xi^2$ histogram and estimates its noise model. Since templates of
lighter binary systems such as \ac{BNS} tend to not couple with non-Gaussian
noise, e.g., \textit{glitches}, most of noise triggers associated with such
low-mass templates are expected to be modeled by the Gaussian noise component.
For each detector, trigger's \ac{SNR} is calculated as square root of a quadrature sum of \ac{SNR}
values given by two orthogonal templates through matched-filtering, each of
which follows the normal distribution. Therefore, the \ac{SNR}-squared follows a
chi-square distribution with two degree-of-freedom regardless of the observed
$\xi^2$ value.

Also, according to the formalism of the upgraded $\rho-\xi^2$ signal model
described in Ref.~\cite{Tsukada:2023}, the expected value of $\xi^2$ in the
\textit{noise} model can be also given by substituting the mismatch factor $k=1$, i.e.,
maximum mismatch between a hypothetical signal and template, into Eq.(40) of Ref.~\cite{Tsukada:2023},
\begin{align}
    \langle\xi^2\rangle=\frac{1}{N}\left\{N+(\rho^2-1)\vec{R}^\dagger\vec{R}\right\}\sim 1,
\end{align}
where $N$ is a auto-correlation length used for $\xi^2$ calculation and
$\vec{R}$ is an auto-correlation function of a given template normalized so that
$R[0]=1$. This approximation holds because an auto-correlation of \ac{CBC}
signals in general sharply peaks at the center where two templates exactly
align, and hence $\vec{R}^\dagger\vec{R} \ll N(R[0])^2=N$. Therefore, 
triggers from the Gaussian noise component are distributed near $\xi^2=1$ and
decaying toward higher \ac{SNR}. Consistently, this behavior is illustrated in \figref{fig:bg_plot}, which shows a $\rho-\xi^2$ noise model for one of the low-mass template bins collected for \ac{LIGO} Hanford detector during the mock data campaign described in Ref.~\cite{mdc_analytics} with the lightblue dashed line as a $\xi^2=1$ contour. In contrast, a tail of the distribution, which extends toward higher \ac{SNR} and $\xi^2$, is known as non-Gaussian noise component.

However, during the online analysis of \ac{O4}, we discovered that the Gaussian
component of the noise model deviates from the expected chi-square
distribution significantly at $\rho>9$ because very few or zero noise trigger
populates such a higher \ac{SNR} regime and the \ac{KDE} applied as a smoothing
process does not complement the lack of triggers sufficiently.
See \figref{fig:snr_pdf} where a 2D $\rho-\xi^2$ noise model is sliced at the $\xi^2=1$ contour and projected onto the $\rho$ dimension. Note that the expected chi-squared distribution shown as a black curve is given by
\begin{align}
    \label{eq:snr_pdf_chi2}
    p(\rho)=\chi^2_{2}(\rho)\frac{\mathrm{d}\rho^2}{\mathrm{d}\rho} \propto 2\rho \mathrm{e}^{-\rho^2/2}.
\end{align}
Although, this disagreement can eventually improve if we keep the analysis
running and collect noise triggers long enough, early phase of the online
analysis is likely to overestimate an event's significance due to this bias and
\nocite{2024GCN.36812....1L}send out a public alert, e.g., S240422ed~\cite{2024GCN.36236....1L}\footnote{The
statistical significance of this candidate was subsequently estimated to be
lower after collecting more noise triggers and the update alert was sent~\cite{2024GCN.36812....1L}}.
Also, we cannot guarantee that this bias
does not exist even in an offline analysis where noise triggers collected
entirely from single (or subset of) observing run are considered. Therefore, the objective here
is to modify the tail of the Gaussian noise component such that it can produce
more accurate noise model even with fewer collected triggers and eventually
prevent a potential false claim of \ac{GW} detections.

To this end, after collecting enough noise triggers, we extrapolate the trigger
counts along a given iso-$\xi^2$ contour using the bulk of the Gaussian component
so that its tail part follows the expected distribution shown in
\eqref{eq:snr_pdf_chi2}. This process is iterated over a range of
$-0.5<\log\xi^2<0.5$ and finally the \ac{KDE} is applied to produce a smoother
distribution. The green curve in \figref{fig:snr_pdf} shows that the noise model
after this extrapolation closely follows the expected distribution in black
curve. Although it can be slightly above the black curve, one can see that it
still provides much more accurate estimate of the event's significance than the
red curve without the extrapolation at $\rho>9$. 

\subsection{New Extinction Model}
A set of candidates undergo two rounds of clustering, the first with other candidates within the template
bin, and the second with candidates from all template bins. These clustering processes alter the \ac{LR} distribution
of candidates. Consequently, the \ac{LR} distribution of the noise triggers which gets used to convert the \ac{LR} of candidates
to FARs also needs to undergo these clustering processes. The new extinction model quantifies the change in 
LR distribution due to clustering, and applies it to the \ac{LR} distribution on noise triggers. To mimic clustering,
this is performed in two steps, the first within the noise triggers of a template bin, and the second with noise
triggers across template bins. Additionally, the new extinction model enforces that any template bin's
contribution to the noise \ac{LR} distribution is proportional to its contribution to the \ac{LR} distribution of
candidates. Both of these features ensure that the FARs assigned to candidates are more accurate
than the old method, i.e., stitching the \ac{LR} distribution of candidates below some \ac{LR} threshold
to the \ac{LR} distribution of noise triggers~\cite{messick2019detecting}.

Let $y(L)$ be the histogram of noise triggers that GstLAL stores in order to estimate the
complementary cumulative distribution function of noise \acp{LR}, $N(L)$.
The two are related in the following way:
\begin{equation}
\label{eq:eq1}
N(L) = A\int_{L}^{\infty} y(L^*) dL^*
\end{equation}
\begin{equation}
\label{eq:eq2}
N_c(L) = A\int_{L}^{\infty} y_c(L^*) dL^*
\end{equation}
where A is some normalization constant, and the subscript $c$ denotes the clustered version of the respective functions.
Equivalently,
\begin{equation}
\label{eq:eq2p5}
y(L) = -\frac{dN(L)}{AdL}
\end{equation}
The new extinction model assumes that the process of obtaining triggers above a certain \ac{LR} threshold $L^*$ in the clustering interval
is a Poisson process, with Poisson rate given by:
\begin{equation}
\label{eq:eq3}
\lambda = c \times N(L^*)
\end{equation}
where $c$ is another normalization constant.
Hence the probability that some trigger with \ac{LR} = $L^*$ survives clustering, i.e. that there
are no triggers with a higher \ac{LR} than $L^*$ in the clustering interval is the Poisson probability for zero events.
\begin{equation}
\label{eq:eq4}
P_{\mathrm{survival}}(L^*) = e^{-\lambda} = e^{-cN(L^*)}
\end{equation}
Hence, we can add the effect of clustering to $y(L)$, by multiplying it with the survival probability.
\begin{equation}
\label{eq:eq5}
y_c(L) = y(L) e^{-cN(L)}
\end{equation}
Substituting \eqref{eq:eq5} in \eqref{eq:eq2}, it follows
\begin{align}
N_c(L) &= A\int_{L}^{\infty} y(L^*) e^{-cN(L^*)} dL^* \nonumber \\
&= - \int_{N(L)}^{N(\infty)} e^{-cN(L^*)} dN(L^*) \nonumber \\
&= \frac{1}{c} [1 - e^{-cN(L)}] 
\end{align}
where $N(\infty)$ = 0. Since GstLAL deals with the complementary cumulative histogram of \acp{LR}, $n(L)$, rather than probability
density, we can convert this to the histogram form by adding another normalization constant. After absorbing relevant constants in $A$ and $c$, we get:
\begin{equation}
\label{eq:eq6}
n_c(L) = A[1 - e^{-cn(L)}]
\end{equation}

In order to find the constants $A$ and $c$, the new extinction model performs a curve fit of $n_c(L)$ to the \ac{LR} histogram of candidates. This is first
done within a template bin, and after all of those clustered bin-specific noise histograms are added together, it is performed across template bins.
The constant $A$ takes care of the relative contributions of template bins to the candidate set, whereas the constant $c$ takes care
of the effect of clustering on the set of candidates.
The curve fitting is performed from the 50th percentile to the 99th percentile of the complementary cumulative histogram of candidate \acp{LR}. This is
empirically known to be a region of well-modeled noise candidates.
The uncertainties in the curve fitting can be calculated from the assumption that a particular realization of the value of $y(L)$ is Poisson 
distributed. Hence,
\begin{align}
\sigma_{y(L)} &= \sqrt{y(L)} \\
\sigma^2_{N(L)} &= \int_{L^*}^{\infty} \sigma^2_{y(L^*)} dL^* \nonumber \\
&= \int_{L^*}^{\infty} y(L^*) dL^* \nonumber \\
\sigma_{N(L)} &= \sqrt{N(L)}
\end{align}

\begin{figure}
\includegraphics[width=\columnwidth]{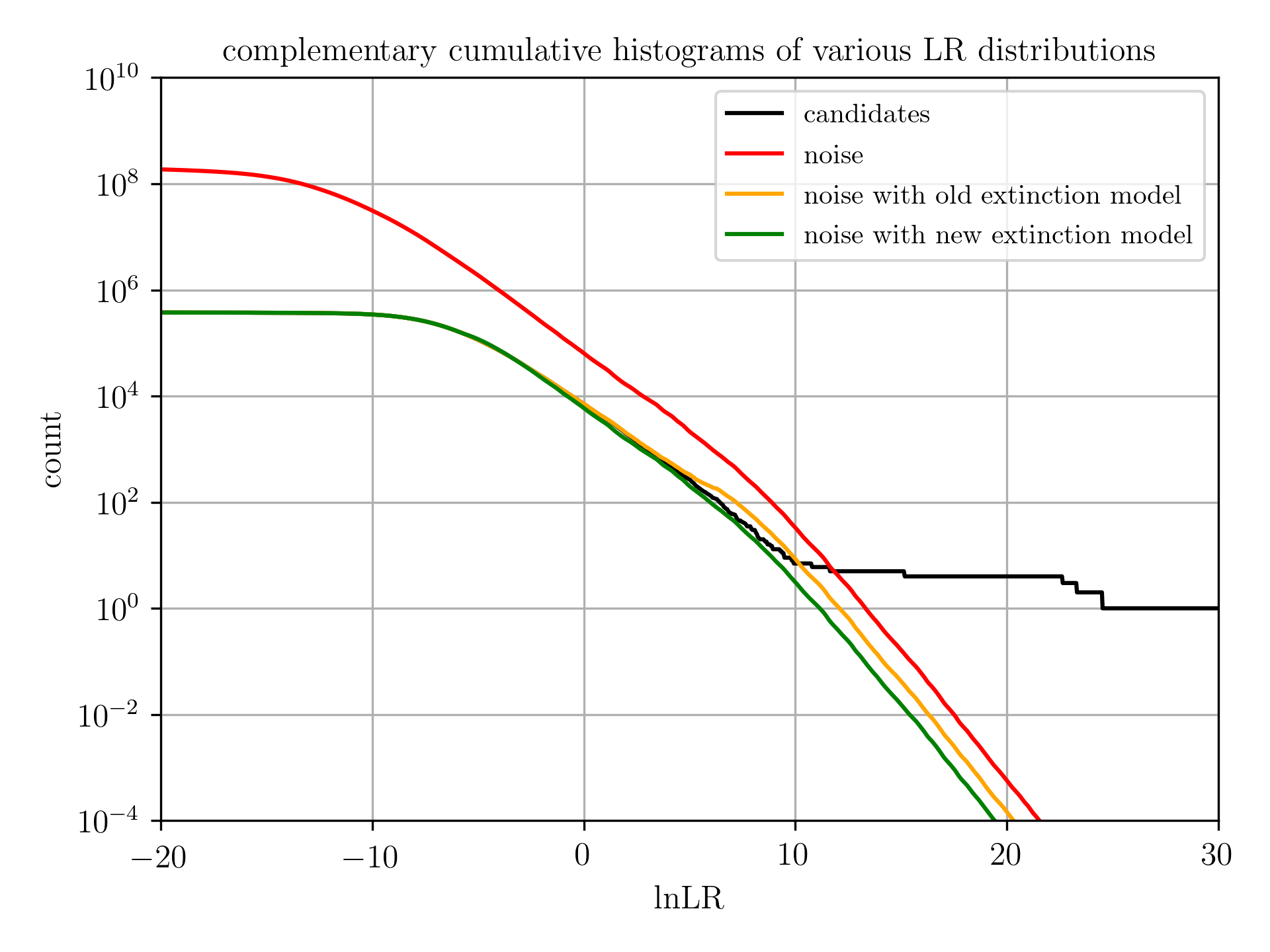}
\caption{\label{fig:extinction}
Effect of applying the new extinction model on the noise \ac{LR} histogram.
The effects of candidate clustering and differing relative contributions of template bins to the candidate set are
modeled by modifying the noise \ac{LR} histogram, n(L) to $A[1 - e^{-cn(L)}]$. The values of A and c are found by
curve fitting the modified noise \ac{LR} histogram to the candidate \ac{LR} histogram. We can see that this process is
effective from the fact that the noise \ac{LR} histogram with the new extinction extinction is very close to the candidate LR
histogram in the noise region (low LR).
}
\end{figure}

The effect of applying the new extinction model on the noise triggers is shown in \figref{fig:extinction}.
We can see that in the noise regime (i.e. 50th percentile to the 99th percentile), the noise \ac{LR} histogram with the new extinction model is 
very close to the candiate histogram. This shows that the new extinction model is successful in applying the effects
of candidate clustering and relative contributions of template bins to the candidate set. By adopting the new
extinction model, we get more accurate FARs, since the noise \ac{LR} distribution used to assign FARs is more accurate than the old method. The new extinction model has also been adopted for GstLAL's online operations in the \ac{O4b}.

\section{Results}
\label{sec:results}

\subsection{Data set}
To test the new methods described in \secref{sec:methodology}, and to serve as a trial run for
GstLAL's \ac{O4} offline results, we set up a GstLAL analysis over 1 chunk of LIGO and Virgo
\ac{O3} data. The data extends for approximately one week, from May 12 19:36:42 UTC 2019 to
May 21 14:45:08 UTC 2019. Since \refref{\cite{joshi2025timesmatchedfiltergravitational}} already tests the online rank feature
and concludes it is equivalent to a traditional offline analysis, we do not test that feature
here, and directly set up an offline analysis over the data instead of running an online 
analysis and setting up an online rank based on that. The offline analysis included a search
with the AllSky bank, and a search with the \ac{IMBH} bank, which were then combined into
a single search using the procedure described in \secref{sec:methodology}, referred to as the ``\ac{O4} analysis'' hereafter. It also includes
the LR improvements, the extrapolation in $\rho-\xi^2$ noise model and new extinction model.
We then compared the results to
GstLAL's \ac{O3} offline results over the same period of data, referred to as the ``\ac{O3} analysis'' hereafter.  Also, note that the \ac{O3} analysis uses the \ac{O3} template bank described in \secref{subsubsection:o3bank}. Both analyses also included
the same injection campaign. The details of the distribution of injections in this campaign
can be found in Ref.~\cite{gwtc-2,LIGO_Virgo_2020}.

\subsection{Candidate lists}
This set of data contains 6 \ac{GW} candidates previously reported in Ref.~\cite{gwtc-2, gwtc-2.1}.
Both the \ac{O4} and \ac{O3} analyses recover all 6 candidates in the list of top 10 candidates, as summarized in \tabref{tab:o4}
and \tabref{tab:o3} respectively.
The \ac{O4} search recovers all 6 confidently with a \ac{FAR} below 1/month, whereas the \ac{O3} search
only recovers 5 of the 6 confidently.

\begin{table*}[]
    \centering
    \begin{tabular}{p{0.5in}p{0.8in}p{0.5in}p{1.25in}p{0.7in}p{0.7in}p{0.5in}p{0.5in}p{0.5in}p{0.5in}}
        \textbf{Rank} & \textbf{FAR} (Hz) & \textbf{SNR} & \textbf{UTC Time} & \textbf{Found\hspace{0.2in}\hspace{0.2in}Detectors} & \textbf{Operating Detectors} & $\mathbf{m_1}$ ($M_{\odot}$) & $\mathbf{m_2}$ ($M_{\odot}$) & $\mathbf{s_{1z}}$ & $\mathbf{s_{2z}}$ \\
        \hline
	1 & $2.69 \times 10^{-38}$ & 24.51 & 2019-05-21 07:43:59 & H1,L1 & H1,L1 & 54.75 & 29.25 & -0.2011 & -0.2011 \\
	2 & $1.46 \times 10^{-20}$ & 13.35 & 2019-05-19 15:35:44 & H1,L1 & H1,L1 & 83.15 & 79.75 & 0.6033 & 0.6033 \\
	3 & $8.60 \times 10^{-19}$ & 14.14 & 2019-05-21 03:02:29 & H1,L1 & H1,L1,V1 & 213.1 & 137.4 & 0.8712 & 0.8712 \\
	4 & $3.72 \times 10^{-16}$ & 12.30 & 2019-05-13 20:54:28 & H1,L1,V1 & H1,L1,V1 & 50.36 & 48.29 & 0.6033 & 0.6033 \\
	5 & $3.56 \times 10^{-12}$ & 10.48 & 2019-05-17 05:51:01 & H1,L1 & H1,L1,V1 & 50.36 & 40.86 & 0.7889 & 0.7889 \\
	6 & $1.65 \times 10^{-10}$ & 10.88 & 2019-05-20 10:35:19 & L1,V1 & H1,L1,V1 & 442.1 & 218.2 & 0.8646 & 0.8646 \\
	7 & $9.71 \times 10^{-9}$ & 8.41 & 2019-05-14 06:54:16 & H1,L1 & H1,L1 & 54.75 & 57.08 & -0.2320 & -0.2320 \\
	8 & $1.25 \times 10^{-8}$ & 13.00 & 2019-05-15 06:27:00 & H1,L1 & H1,L1 & 543.3 & 63.58 & 0.2638 & 0.2638 \\
	9 & $1.46 \times 10^{-8}$ & 13.67 & 2019-05-14 04:01:29 & H1,L1 & H1,L1 & 414.9 & 144.7 & 0.8778 & 0.8778 \\
	10 & $1.88 \times 10^{-8}$ & 8.39 & 2019-05-18 20:43:46 & L1,V1 & H1,L1,V1 & 414.9 & 218.2 & 0.8646 & 0.8646 \\
    \end{tabular}
    \caption{The candidate list for the offline search using GstLAL's \ac{O4} methods. The search is run
over a week of \ac{O3} data. The first, second, third, fourth, fifth, and seventh candidates in this list correspond to
the \acp{GW} GW190521\_074359, GW190519\_153544, GW190521\_030229, GW190513\_205428, GW190517\_055101, and GW190514\_065416, previously
reported in Ref.~\cite{gwtc-2, gwtc-2.1}. The remaining 4 candidates all arise from the \ac{IMBH} search, and have not been previously reported. Given the uncertainty surrounding
the \ac{IMBH} space of \ac{GW} signals, we make no claims regarding the origin of these 4 candidates}
    \label{tab:o4}
\end{table*}

\begin{table*}[]
    \centering
    \begin{tabular}{p{0.5in}p{0.8in}p{0.5in}p{1.25in}p{0.7in}p{0.7in}p{0.5in}p{0.5in}p{0.5in}p{0.5in}}
	\textbf{Rank} & \textbf{FAR} (Hz) & \textbf{SNR} & \textbf{UTC Time} & \textbf{Found\hspace{0.2in}\hspace{0.2in}Detectors} & \textbf{Operating Detectors} & $\mathbf{m_1}$ ($M_{\odot}$) & $\mathbf{m_2}$ ($M_{\odot}$) & $\mathbf{s_{1z}}$ & $\mathbf{s_{2z}}$ \\
        \hline
	1 & $1.04 \times 10^{-43}$ & 24.43 & 2019-05-21 07:43:59 & H1,L1 & H1,L1 & 56.33 & 37.81 & -0.0015 & 0.4993 \\
	2 & $4.33 \times 10^{-21}$ & 11.68 & 2019-05-13 20:54:28 & H1,L1 & H1,L1 & 60.74 & 32.32 & 0.8530 & -0.1904 \\
	3 & $1.58 \times 10^{-18}$ & 13.03 & 2019-05-19 15:35:44 & H1,L1 & H1,L1 & 103.40 & 11.61 & 0.0925 & -0.9660 \\
	4 & $2.35 \times 10^{-11}$ & 10.36 & 2019-05-17 05:51:01 & H1,L1,V1 & H1,L1,V1 & 49.39 & 39.63 & 0.6712 & 0.8675 \\
	5 & $2.80 \times 10^{-10}$ & 12.94 & 2019-05-21 03:02:29 & H1,L1,V1 & H1,L1,V1 & 139.70 & 2.92 & -0.2471 & -0.0434 \\
	6 & $3.23 \times 10^{-9}$ & 8.42 & 2019-05-14 23:48:57 & L1,V1 & L1,V1 & 2.34 & 1.24 & -0.0130 & 0.0336 \\
	7 & $3.16 \times 10^{-7}$ & 8.93 & 2019-05-20 06:39:01 & H1,L1 & H1,L1 & 4.94 & 1.10 & 0.3486 & -0.0499 \\
	8 & $2.70 \times 10^{-6}$ & 8.33 & 2019-05-14 06:54:16 & H1,L1 & H1,L1 & 53.48 & 53.48 & -0.5578 & -0.5578 \\
	9 & $3.02 \times 10^{-6}$ & 9.63 & 2019-05-20 20:45:36 & H1,V1 & H1,L1,V1 & 21.42 & 18.44 & 0.6018 & 0.4931 \\
	10 & $3.85 \times 10^{-6}$ & 8.92 & 2019-05-17 14:43:16 & H1,L1 & H1,L1 & 19.96 & 1.24 & -0.9457 & 0.0449 \\
    \end{tabular}
    \caption{The candidate list for the offline search using GstLAL's \ac{O3} methods. The search is run
over a week of \ac{O3} data. The first, second, third, fourth, fifth, and eighth candidates in this list correspond to the \acp{GW} GW190521\_074359,
GW190513\_205428, GW190519\_153544, GW190517\_055101, GW190521\_030229, and GW190514\_065416,
previously reported in Ref.~\cite{gwtc-2, gwtc-2.1}.}
    \label{tab:o3}
\end{table*}

\subsection{Sensitivity comparison}
\begin{figure}
\includegraphics[width=\columnwidth]{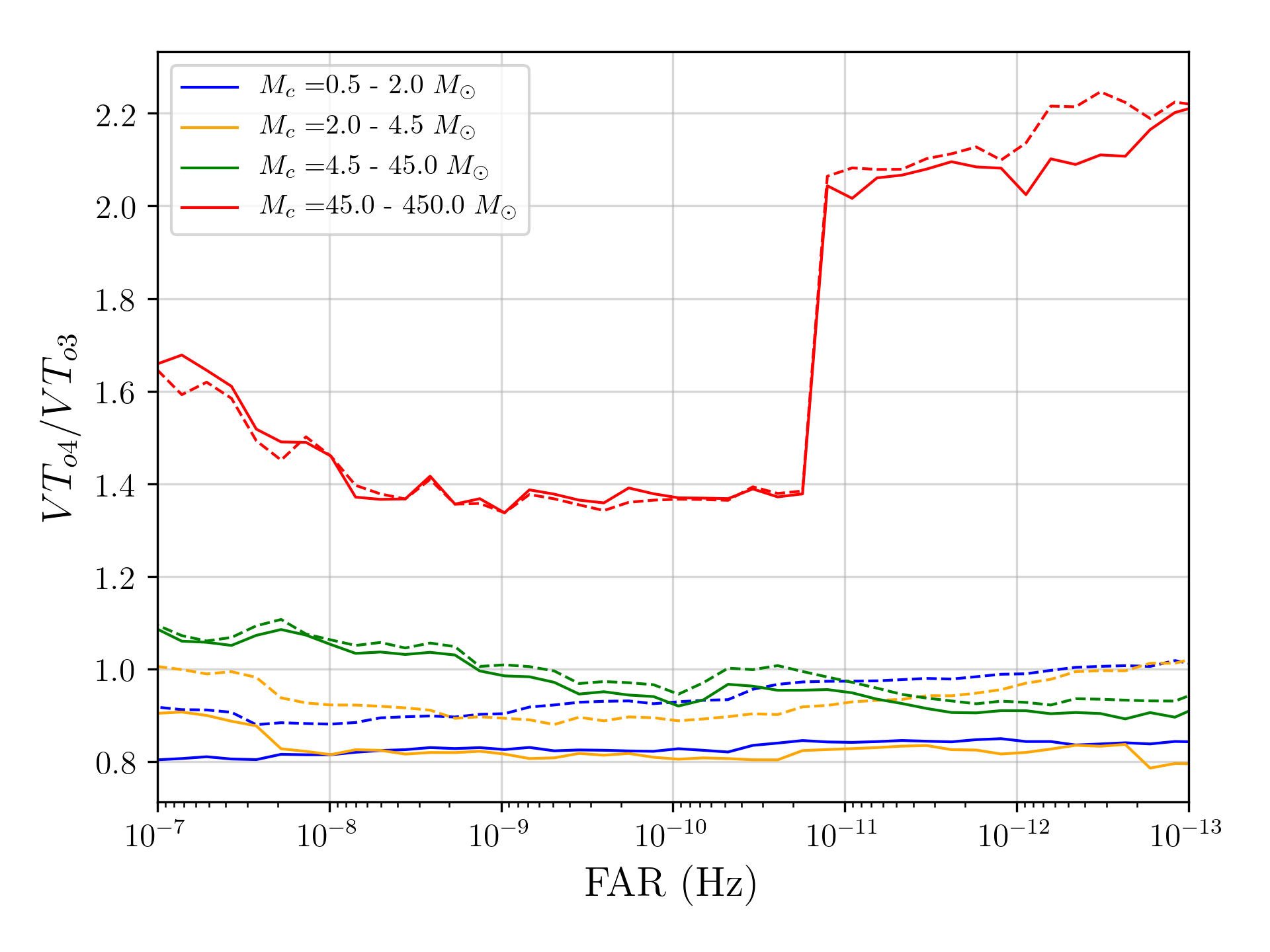}
\caption{\label{fig:vt_ratio}
\textit{VT} ratios of the \ac{O4} search with the \ac{O3} search.
The solid lines represent the full \ac{O4} search as compared to the \ac{O3} search,
whereas the dashed lines represent the \ac{O4} search without the extrapolation in $\rho-\xi^2$ noise model as compared to the \ac{O3} search.
Since the extrapolation removes false positives from the candidates, it corrects previous mis-estimations of the \textit{VT}
done in the \ac{O3} search, superficially lowering the \ac{O4} \textit{VT}. The \textit{VT} ratios of the two
lowest mass bins are slightly less than 1. This is because the \ac{O3} temlpate bank sampled the lower mass parameter space
with a higher minimum match (0.99) as compared to \ac{O4} (0.97). Additionally, we see that the highest mass bin has a 50\%
- 100\% increase in sensitivity, arising from the LR improvements and the additional \ac{IMBH} search done in \ac{O4}.
}
\end{figure}

To compare the sensitivities of the \ac{O4} and \ac{O3} analyses, we use the sensitive volume-time, or \textit{VT} of the
two analyses as a metric. It represents the volume of 4-dimentional space-time where the search can typically identify GW signals. Since the
times analyzed by both the \ac{O4} and \ac{O3} analyses is exactly the same, the \textit{VT}s of the two analyses
are a measure of their relative sensitivities. Furthermore, the \textit{VT}s can be calculated for different mass ranges
in order to estimate the relative sensitivities to different source classes.
A plot of the ratio of the \textit{VT} of the \ac{O4} analysis with
and without the extrapolation in $\rho-\xi^2$ noise model, with the \ac{O3} analysis is shown in \figref{fig:vt_ratio}.

\subsection{Performance of individual features}

\subsubsection{IMBH analysis}
\begin{figure}
\includegraphics[width=\columnwidth]{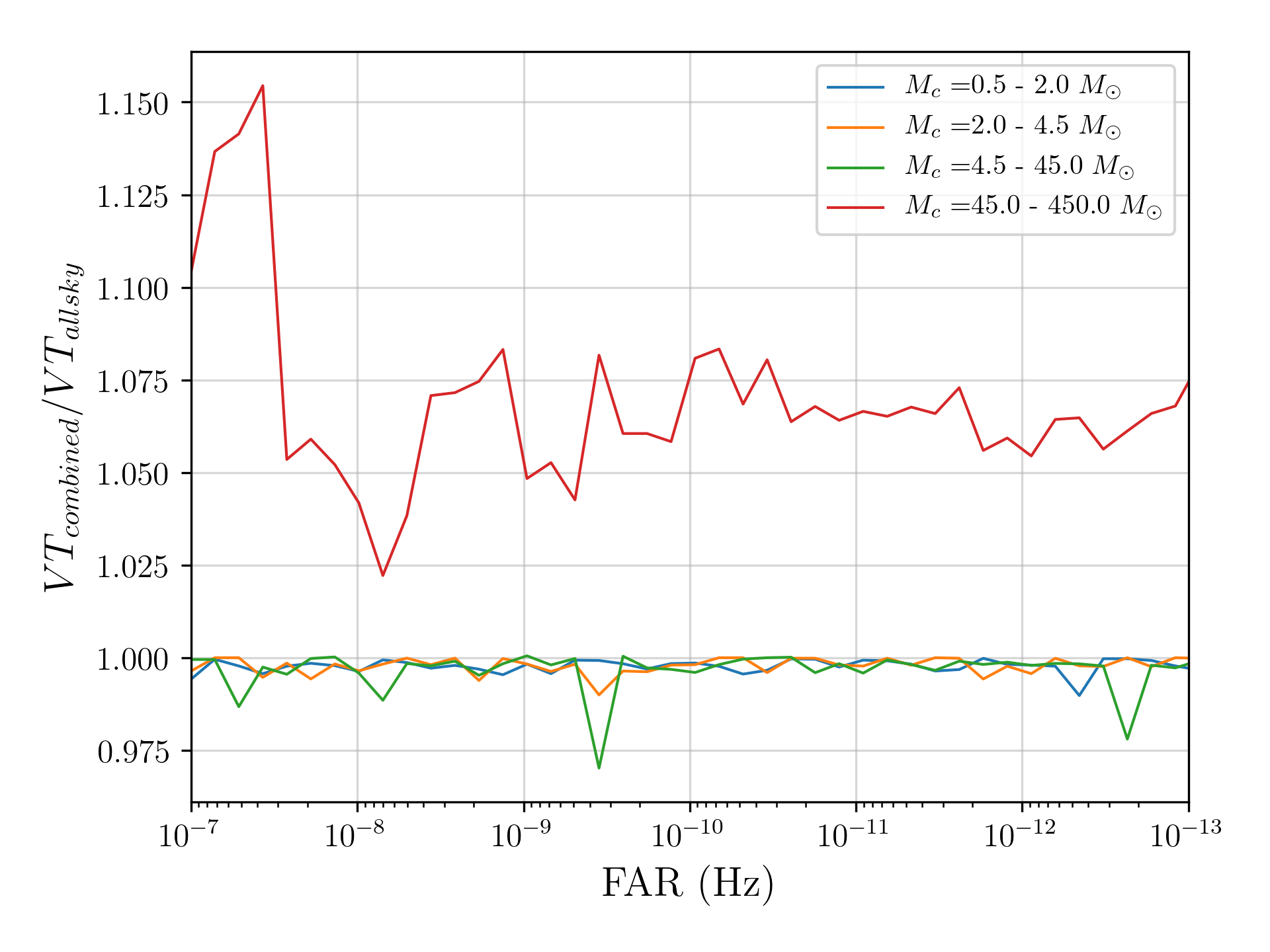}
\caption{\label{fig:vt_ratio_imbh}
This plot shows the ratio of \textit{VT} of the \ac{O4} combined AllSky+IMBH search
to that of the \ac{O4} AllSky search, and it shows us how much sensitivity we gain
by combining the IMBH and AllSky searches, as compared to just the AllSky search.
As expected the sensitivities of the 3 lowest mass bins are unchanged, since the IMBH
bank is not sensitive in that region. The \textit{VT} of the IMBH bin increases
by 6\% - 7\%.
}
\end{figure}

Here, we assess how much sensitivity we gain by combining the \ac{IMBH} search with the AllSky search, compared to
just the sensitivity of the AllSky search. The \textit{VT} ratio of the \ac{O4} combined AllSky+IMBH search to that
of the \textit{VT} of the \ac{O4} AllSky search is shown in \figref{fig:vt_ratio_imbh}.
In addition to the 6\% - 7\% sensitivity improvement in the \ac{IMBH} space as shown in \figref{fig:vt_ratio_imbh},
by adding the \ac{IMBH} search to the AllSky one, the combined analysis becomes sensitive to a new parameter space,
and has the ability to recover \ac{GW} signals arising from intermediate-mass black hole mergers, which the AllSky search
did not have by itself. This source class is particularly interesting and by detecting \acp{GW} from such mergers,
we enable new scientific results~\cite{abbott2022search}.

\subsubsection{Extrapolation in $\rho-\xi^2$ noise model}
\begin{figure}
\includegraphics[width=\columnwidth]{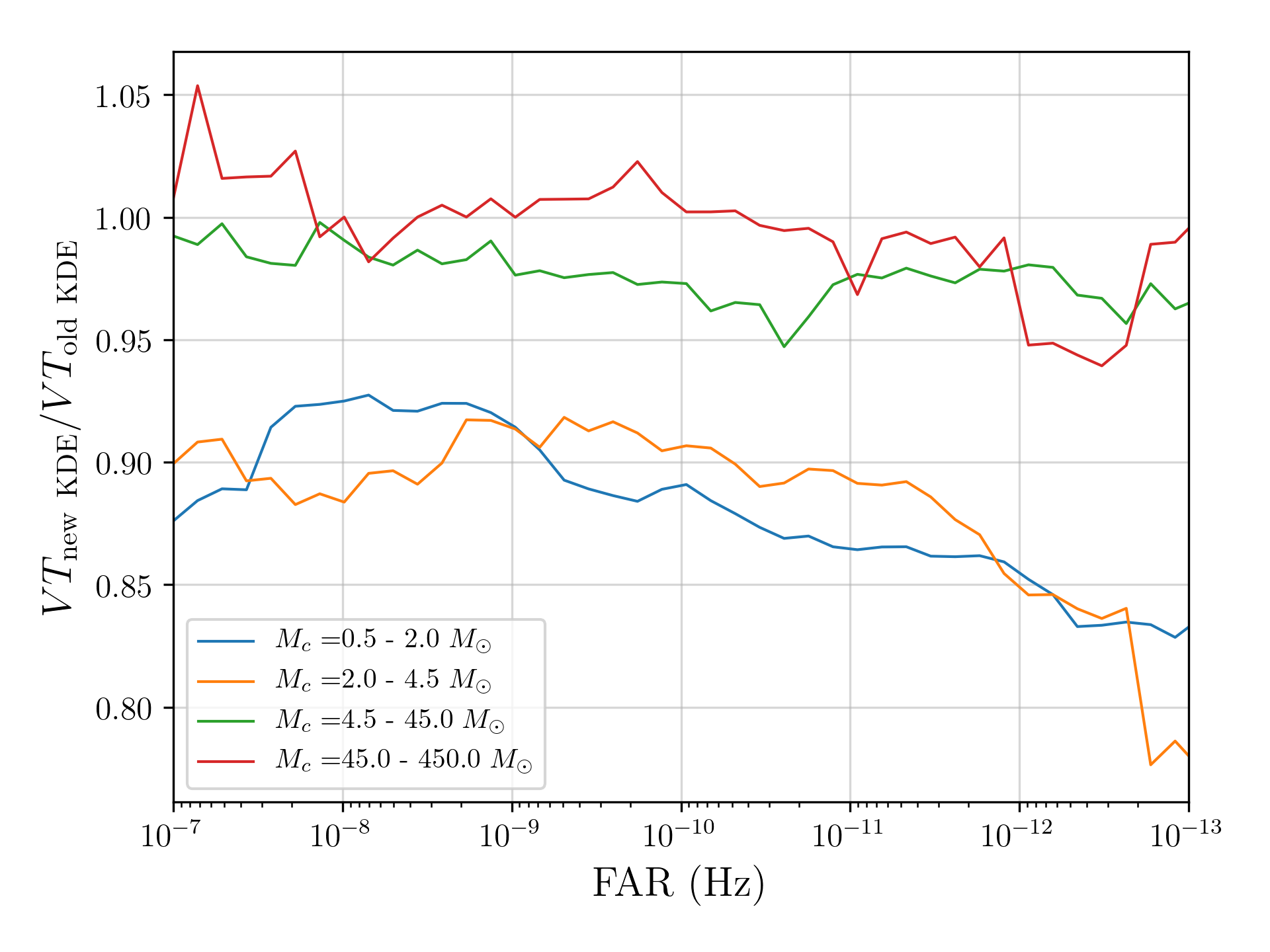}
\caption{\label{fig:vt_ratio_kde}
\textit{VT} ratio of the \ac{O4} search with the noise
model extrapolation as compared to without. The extrapolation removes false
positives from the set of candidates, and in the process also downweights a
small amount of real \ac{GW} signals. The former effect corrects the
overestimate of \textit{VT} that happens without the extrapolation, whereas the
latter decreases the real sensitivity of the search. Both these effects
contribute to the lower than 1 \textit{VT} ratio seen here. This effect is
mostly seen in the BNS ans NSBH regions, whose \textit{VT} goes down by 10\% -
15\%. The BBH and IMBH regions are mostly unaffected.
}
\end{figure}

As mentioned previously, the extrapolation in $\rho-\xi^2$ noise model removes false positives in \ac{GW} candidates.
This implies that real
\ac{GW} signals can also be downranked accordingly and not be recovered as significantly as
it would be without the extrapolation. To this end, we compare the $VT$s
between with and without the extrapolation to assess the signal recovery performance. \figref{fig:vt_ratio_kde} shows the
$VT$ of the O4 analysis with the noise model extrapolation as a
function of the recovered \ac{FAR} for each category of injections, being
relative to that for without the extrapolation. Note that the sensitivity can
decrease up to 15\% for two lowest mass categories of the injections, e.g., blue
and orange curves. This is because, as mentioned above, the noise model for
lower mass templates tends to be dominated by the Gaussian component and hence
the extrapolation is likely to be more impactful. In contrast, for heavier BBH
templates, the non-Gaussian noise component overwhelms the Gaussian one and
makes small changes at its tail part negligible. We also emphasize that this
decrease in the search sensitivity does \textit{not} necessarily indicate the
lowered performance overall, but rather that the signal recovery without the
extrapolation is overestimated and risks potential false positives.

\subsubsection{Reliability of results}
The new extinction model and LR improvements ensure the \acp{FAR} of candidates are more accurate, and the extrapolation in $\rho-\xi^2$ noise model reduces the risk of false positives in the candidate set. All of these contribute
to the \ac{O4} results being more reliable then before. This effect is difficult to measure
since wrongly assigned \acp{FAR} and false positives are rare. One possible metric is to see how
many candidates in the injection campaign with a \ac{FAR} below some threshold do not correspond
to an injected \ac{GW} signal. With a \ac{FAR} threshold of one per day, this number for the
\ac{O3} search is 53, representing 0.07\% of all candidates. The corresponding number for the \ac{O4} search
is 28, representing 0.03\% of all candidates. For a higher \ac{FAR} threshold of one per hour.
the numbers for the \ac{O3} and \ac{O4} searches is 276 (0.35\%) and 132 (0.15\%). This shows that the 
new \ac{O4} methods are successful in lowering the number of false positives.

The \ac{O4} search includes the \ac{IMBH} search. This parameter space is known to be particularly
vulnerable to loud noise transients or non-gaussian transient ``glitches"~\cite{Ghosh_2024, PhysRevD.105.103021}. If we only look at the 
\ac{O4} AllSky search, for a more direct comparison with \ac{O3}, the numbers for the one-per-day
and one-per-hour \ac{FAR} thresholds are 23 (0.03\%), and 110 (0.13\%), showing an
even higher efficacy of removing false positives.

\section{Conclusion}
\label{sec:conclusion}
In this work, we first gave an overview of the GstLAL offline analysis.
This included desriptions of the GstLAL AllSky and \ac{IMBH} template banks used in \ac{O4}.
We later described new methods in the GstLAL offline analysis introduced in the lead up to
and during \ac{O4}. These new methods are all used to obtain GstLAL's offline 
results for \ac{O4}.

The new methods include a way to outsource everything up to the matched filtering
stage to an online analysis that had previously run on the same data. This, along with improvements
to modularity of the workflow and reusability of results means that \ac{GW} data only needs
to be matched filtered with a given template bank only once during an observing run. The results
of any such matched filtering, whether online or offline, and with any template bank can be
combined into a single cohesive set of results.

The new methods also include a new \ac{IMBH} offline search. This search only allows candidates with multiple contributing detectors, in order to make the search well-behaved and functional. A new method
to combine the AllSky and \ac{IMBH} searches was also introduced. It assigns a weight to each
search by which the FARs of their candidates, calculated independently for each search, are scaled up.
Afterwards, the set of candidates of the two searches can be combined in to single set of candidates.

Finally, we also discussed some changes to the LR calculation, introduced the extrapolation in $\rho-\xi^2$ noise model, as well as a
new extinction model which enables a better estimation of the LR statistics of noise triggers.

To test these new methods, we set up an offline analysis over one week of O3 data, and compared the
results with those obtained from the same data by GstLAL during O3. Both searches recovered the 
6 previously reported GWs in the data in their list of top 10 candidates. The O4 search was able to do so confidently 
for all 6, whereas the O3 search managed it only for 5. We found that the sensitivity of the O4 search in
the \ac{IMBH} space increased by 50\% - 100\%. The \textit{VT} in the BNS and NSBH space went down by 10\% -
20\% as a result of the extrapolation in $\rho-\xi^2$ noise model, which reduces false positives and downweighting a small number of real \ac{GW}
signals accordingly. Without the new KDE, those \textit{VT}s are very close to the O3 ones.

The new methods described in this work variously increase the sensitivity, reliability, or reusability
of the GstLAL offline results. As a result, we expect GstLAL's offline results to significantly
contribute to the \ac{LVK}'s scientific results in \ac{O4}.

\begin{acknowledgments}
This research has made use of data, software and/or web tools obtained from the
Gravitational Wave Open Science Center (https://www.gw-openscience.org/ ), a
service of \ac{LIGO} Laboratory, the \ac{LSC} and the Virgo
Collaboration.  
We especially made heavy use of the \ac{LVK} Algorithm
Library. 
\ac{LIGO} was constructed by the California Institute of Technology and the 
Massachusetts Institute of Technology with funding from the United States 
National Science Foundation (NSF) and operates under cooperative agreements 
PHYS-$0757058$ and PHY-$0823459$.
In addition, the Science and Technology Facilities Council (STFC) of the United 
Kingdom, the Max-Planck-Society (MPS), and the State of Niedersachsen/Germany 
supported the construction of \ac{aLIGO} and construction and operation of the 
GEO600 detector. 
Additional support for \ac{aLIGO} was provided by the Australian Research Council.  
Virgo is funded, through the European Gravitational Observatory (EGO), by the 
French Centre National de Recherche Scientifique (CNRS), the Italian Istituto 
Nazionale di Fisica Nucleare (INFN) and the Dutch Nikhef, with contributions by 
institutions from Belgium, Germany, Greece, Hungary, Ireland, Japan, Monaco, 
Poland, Portugal, Spain.

The authors are grateful for computational resources provided by 
the \ac{LIGO} Lab culster at the \ac{LIGO} Laboratory and supported by 
PHY-$0757058$ and PHY$-0823459$, the Pennsylvania State University's Institute 
for Computational and Data Sciences gravitational-wave cluster, 
and supported by 
OAC-$2103662$, PHY-$2308881$, PHY-$2011865$, OAC-$2201445$, OAC-$2018299$, 
and PHY-$2207728$.  
The authors thank Thomas Dent (and the LVK CBC AllSky group) for sharing initial investigations on the Gaussian noise component of the $\rho-\xi^2$ noise model discussed in \secref{subsec:rank_stats}.
The authors also thank Reed Essick (and the entire LVK Rates and Populations group) for developing the injection
sets used in the calculation of the weights for the AllSky and IMBH analyses.
LT acknowledges NASA 80NSSC23M0104 and the Nevada Center for Astrophysics for support.
CH Acknowledges generous support from the Eberly College of Science, the 
Department of Physics, the Institute for Gravitation and the Cosmos, the 
Institute for Computational and Data Sciences, and the Freed Early Career Professorship.
US and SS acknowledge support from NSF PHY-2409714.

\end{acknowledgments}


\clearpage

\bibliography{references}

\end{document}